\documentclass[journal]{IEEEtran}
\usepackage{cite}

\ifCLASSINFOpdf
  \usepackage[pdftex]{graphicx}
  \graphicspath{{./fig/}}
  \DeclareGraphicsExtensions{.pdf, .png}
\else
\fi
\ifCLASSOPTIONcompsoc
 \usepackage[caption=false,font=normalsize,labelfont=sf,textfont=sf]{subfig}
\else
 \usepackage[caption=false,font=footnotesize]{subfig}
\fi
\usepackage{algorithm}
\usepackage{algpseudocode}
\usepackage{amsfonts}
\usepackage{color}
\usepackage{footnote}
\usepackage[inline]{enumitem}
\usepackage{mathtools}
\usepackage{textcomp, gensymb}
\makesavenoteenv{tabular}

\DeclareMathOperator*{\minimize}{minimize}
\DeclareMathOperator*{\prox}{prox}

\DeclareMathOperator*{\T}{T}
\newcommand{\bbR}{\mathbb R}
\newcommand{\bfb}{\mathbf b}
\newcommand{\bfn}{\mathbf n}
\newcommand{\bfu}{\mathbf u}
\newcommand{\bfw}{\mathbf w}
\newcommand{\bfx}{\mathbf x}
\newcommand{\bfy}{\mathbf y}
\newcommand{\bfz}{\mathbf z}
\newcommand{\bftx}{\mathbf{\tilde x}}
\newcommand{\bfty}{\mathbf{\tilde y}}
\newcommand{\bftz}{\mathbf{\tilde z}}
\newcommand{\bfA}{\mathbf A}
\newcommand{\Let}[2]{$#1 \leftarrow #2$}

\begin{document}
%
\title{Spaceborne Staring Spotlight SAR Tomography---A First Demonstration with TerraSAR-X}
%
%
%

\author{Nan~Ge,
        Fernando~Rodriguez~Gonzalez,
        Yuanyuan~Wang,~\IEEEmembership{Member,~IEEE,}
        Yilei~Shi,
        Xiao~Xiang~Zhu,~\IEEEmembership{Senior~Member,~IEEE}
\thanks{This work is jointly supported by Helmholtz Association under the framework of the Young Investigators Group ``SiPEO'' (VH-NG-1018, www.sipeo.bgu.tum.de), and the European Research Council (ERC) under the European Union's Horizon 2020 research and innovation program (ERC-2016-StG-714087, So2Sat). \emph{Corresponding author: Xiao Xiang Zhu.}}

\thanks{Nan Ge and Fernando Rodriguez Gonzalez are with the Remote Sensing Technology Institute, German Aerospace Center (DLR), 82234 Wessling, Germany (e-mail: Nan.Ge@dlr.de; Fernando.RodriguezGonzalez@dlr.de).}
\thanks{Yuanyuan Wang is with Signal Processing in Earth Observation, Technical University of Munich, 80333 Munich, Germany (e-mail: wang@bv.tum.de).}
\thanks{Yilei Shi is with Remote Sensing Technology, Technical University of Munich, 80333 Munich, Germany (e-mail: yilei.shi@tum.de).}
\thanks{Xiao Xiang Zhu is with the Remote Sensing Technology Institute, German Aerospace Center (DLR), 82234 Wessling, Germany, and also with Signal Processing in Earth Observation, Technical University of Munich, 80333 Munich, Germany (e-mail: xiao.zhu@dlr.de).}
}
\maketitle

\begin{abstract}
\textit{This is a preprint. To read the final version please visit IEEE  XPlore.}

With the objective of exploiting hardware capabilities and preparing the ground for the next-generation X-band synthetic aperture radar (SAR) missions, TerraSAR-X and TanDEM-X are now able to operate in staring spotlight mode, which is characterized by an increased azimuth resolution of approximately $0.24$~m compared to $1.1$~m of the conventional sliding spotlight mode. In this paper, we demonstrate for the first time its potential for SAR tomography. To this end, we tailored our interferometric and tomographic processors for the distinctive features of the staring spotlight mode, which will be analyzed accordingly. By means of its higher spatial resolution, the staring spotlight mode will not only lead to a denser point cloud, but also to more accurate height estimates due to the higher signal-to-clutter ratio. As a result of a first comparison between sliding and staring spotlight TomoSAR, the following were observed:
\begin{enumerate*}[label=\arabic*)]
\item the density of the \emph{staring} spotlight point cloud is approximately \(5.1\)--\(5.5\) times as high;
\item the relative height accuracy of the \emph{staring} spotlight point cloud is approximately \(1.7\) times as high.
\end{enumerate*}
\end{abstract}


%
\IEEEpeerreviewmaketitle

\section{Introduction} \label{sec:1}
%
%
%
%

 

\IEEEPARstart{T}{erraSAR-X} and TanDEM-X, the twin German satellites of almost identical build, have been delivering high-resolution X-band synthetic aperture radar (SAR) images since their launch in 2007 and 2010, respectively.  Among civil SAR satellites, their unprecedented high spatial resolution in meter range and relatively short revisit time of $11$~days opened up new applications of spaceborne SAR interferometry (InSAR). As a benchmark of medium-resolution spaceborne SAR sensors, a resolution cell in an ENVISAT ASAR stripmap product of the size $6$-by-$9$~m$^2$ (azimuth-by-range) is resolved by approximately $5$-by-$15$~pixels in a high-resolution sliding spotlight image of TerraSAR-X with $300$ MHz range bandwidth \cite{ein:09}. Particularly in urban areas, this meter-level resolution provides the possibility of revealing detailed information in terms of geolocation and motion of single man-made objects. Adaptations of advanced time series analysis methods, such as persistent scatterer interferometry (PSI) and SAR tomography (TomoSAR), to sliding spotlight datasets showed promising results, see, for example, \cite{ger:10, con:12, zhu:10a, rea:11}.

In order to fully exploit the capabilities of TerraSAR-X\footnote{In the following TerraSAR-X is referred to as the monostatic constellation of TerraSAR-X and TanDEM-X, i.e., SAR instrument is activated on either TerraSAR-X or TanDEM-X but not both.} and to prepare for the next-generation X-band SAR satellite missions, e.g., HRWS \cite{bar:16}, the TerraSAR-X staring spotlight mode was conceptualized and consequently operationalized \cite{mit:14, pra:14}. Compared to the high-resolution sliding spotlight mode, the SAR sensor in staring spotlight mode employs a larger squint angle range to achieve a better azimuth resolution of approximately $0.24$~m. As a result, the same ENVISAT ASAR stripmap pixel, as mentioned in the previous paragraph, is represented by $25$-by-$15$ pixels in a staring spotlight image. The advantages of increased (azimuth) resolution for urban areas are at least two-fold:
\begin{enumerate*}[label=\arabic*)]
\item it is more likely for point-like targets with similar azimuth-range coordinates to appear in different resolution cells, thus densifying the 4-D point cloud;
\item point-like targets stand out more prominently from clutter, which leads to higher signal-to-clutter ratio (SCR).
\end{enumerate*}
These factors favor PSI and TomoSAR in different ways. While the former increases the amount of information of particularly single man-made objects, the latter provides a better lower bound on the variance of height estimates \cite{bam:09}.

Although it seems encouraging to adapt and apply TomoSAR to staring spotlight datasets, yet to the best of our knowledge there has not been any published result. A lack of datasets could be one reason. On the other hand, several considerations regarding staring spotlight mode need to be taken into account during InSAR processing, which might also hinder such an application. By means of this paper, we intend to show that staring spotlight datasets are indeed suitable for TomoSAR. Based on a sufficient number of acquisitions, our first results on the scales of a city and of individual infrastructures are demonstrated to provide an argument in favor of this statement. We also perform a preliminary comparison between sliding and staring spotlight TomoSAR by using a limited number of datasets in both modes.

The remainder of this paper is organized as follows. Section~\ref{sec:2} explains the TerraSAR-X staring spotlight mode and its related InSAR processing aspects. The principles of TomoSAR are briefly revisited in section~\ref{sec:3}, where several technical adaptations are elucidated as well. Section~\ref{sec:4} comprises our first results with an interferometric stack of Washington, D.C.\ and some interpretations thereof. In section~\ref{sec:5}, a preliminary comparison of sliding and staring spotlight TomoSAR is made based on a small number of images. Conclusions are drawn and future work is proposed in section~\ref{sec:6}. The appendix clarifies the structure of the TerraSAR-X annotation component containing a $3$-by-$3$ grid of Doppler centroid in focused image time, which could be used to avoid complex time conversion.

\section{TerraSAR-X Staring Spotlight Interferometry} \label{sec:2}

\begin{figure}[!tp]
\centering
\includegraphics[width=0.48\textwidth]{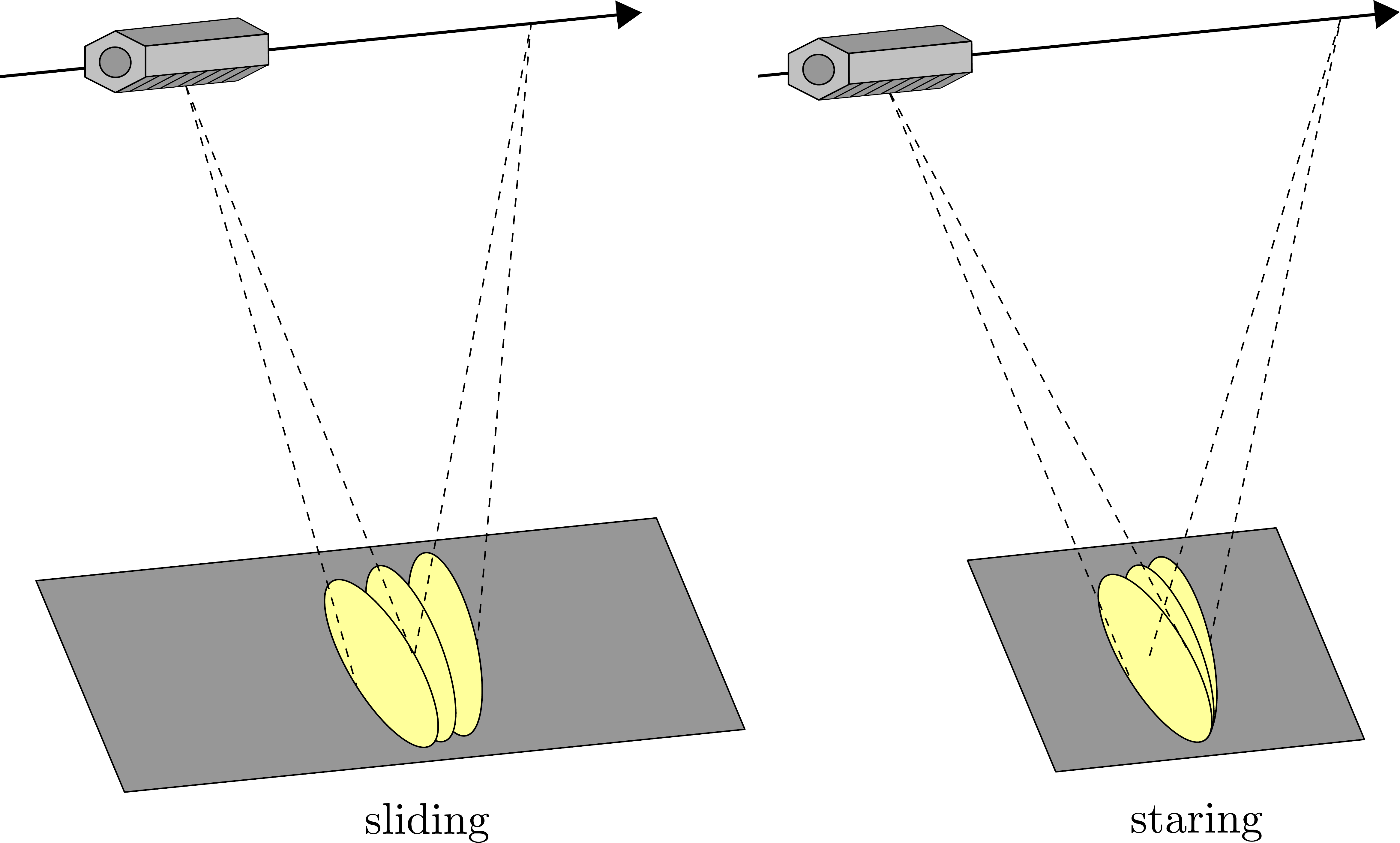}
\caption{TerraSAR-X sliding (left) and staring (right) spotlight imaging geometry. Modified from \cite{ein:09}.} \label{fig:1}
\end{figure}

\begin{figure}[!tp]
\centering
\includegraphics[width=0.35\textwidth]{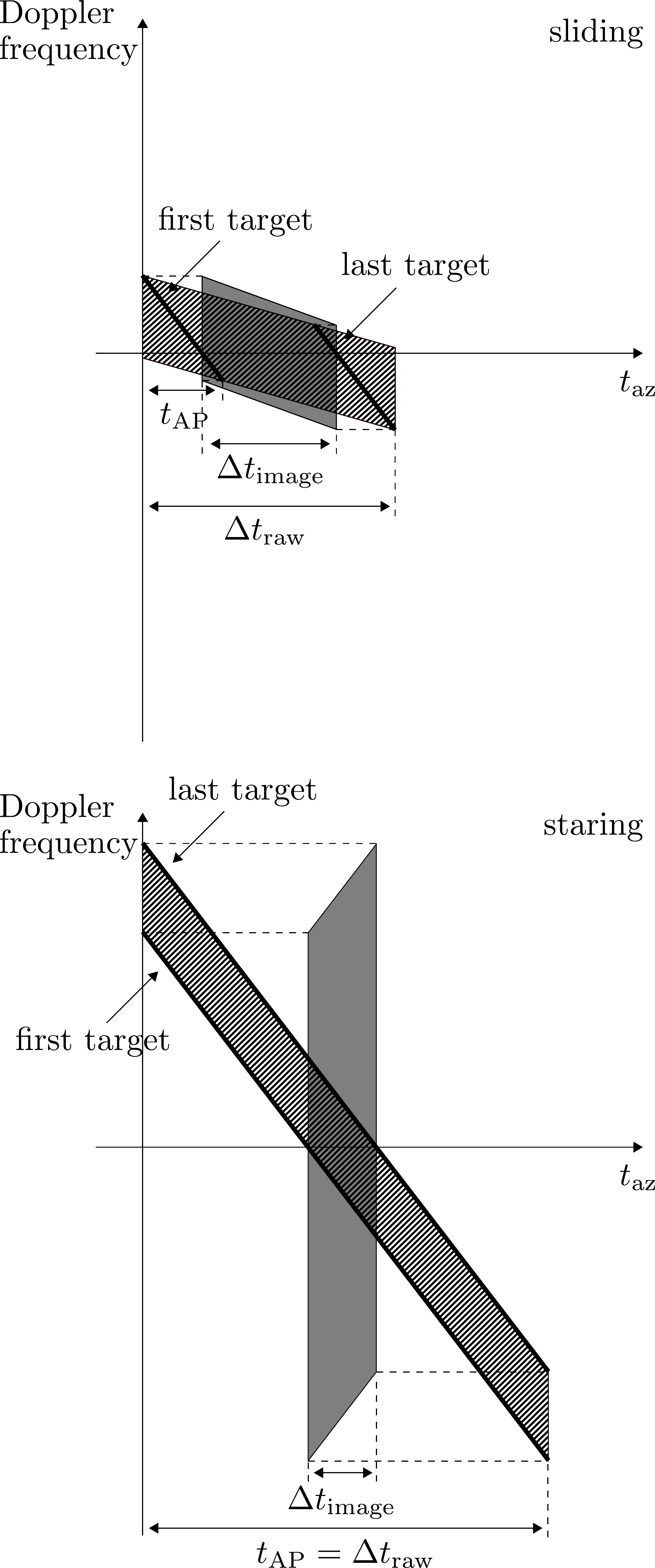}
\caption{Time-variant Doppler spectra of SAR raw data (//) with time span $\Delta t_{\text{raw}}$, and of focused image (shaded) with time span $\Delta t_{\text{image}}$ in the sliding (top, modified from \cite{ein:09}) and staring (bottom) spotlight modes. Bold line segments denote the targets at the start and stop azimuth time ($t_{\text{az}}$) in the focused image, respectively. Both targets are illuminated with time $t_{\text{AP}}$ and their zero-crossings define $\Delta t_{\text{image}}$. In the staring spotlight mode, $t_{\text{AP}}$ is set to equal $\Delta t_{\text{raw}}$ in order to increase the azimuth resolution, which comes at the expense of significantly shorter $\Delta t_{\text{image}}$.} \label{fig:2}
\end{figure}

In spotlight mode, the SAR sensor steers the azimuth beam forth and back in order to increase the illumination (or aperture) time $t_{\text{AP}}$ of a target, as illustrated in Fig.~\ref{fig:1}. As a side effect, the Doppler centroid frequency undergoes a negative drift in azimuth time $t_{\text{az}}$ of the raw data (see Fig.~\ref{fig:2}). The beam sweep rate is a trade-off between azimuth resolution and spatial extent. In the TerraSAR-X sliding spotlight mode, the azimuth beam is swept at a moderate rate with a squint angle range up to $\pm 0.75\degree$ \cite{ein:13}, while in the staring spotlight mode the azimuth beam is steered exactly towards a reference ground target as satellite proceeds. In other words, the beam sweep rate is configured to match the frequency modulation (FM) rate of the reference target, which enables longer azimuth illumination time. To be more specific, the acquisition squint angle range is restricted to approximately $\pm 2.2\degree$ due to antenna azimuth grating lobe \cite{mit:14}. As a consequence, $t_{\text{AP}}$ is, in the ideal case, equal to the azimuth time span of the raw data $\Delta t_{\text{raw}}$. This leads to a maximized azimuth resolution, which is limited by the product of $t_{\text{AP}}$ and the FM rate \cite{ein:09}. This improved azimuth resolution comes, however, at the expense of a reduced azimuth scene extent, i.e., the azimuth time span of a focused image $\Delta t_{\text{image}}$ in staring spotlight mode is significantly shorter. Naturally, the intrinsic range bandwidth imposes a ceiling on the slant range resolution, which is normally solely enhanced by a hardware upgrade. Tab.~\ref{tab:1} lists as an example the parameters of a TerraSAR-X staring spotlight acquisition of Washington, D.C.

\begin{table}[!tp]
\renewcommand{\arraystretch}{1.3}
\caption{Exemplary Parameters of a TerraSAR-X staring spotlight acquisition of Washington, D.C.\ (values are rounded)}
\label{tab:1}
\centering
\begin{tabular}{l c}
\hline
Incidence angle at scene center & $41\degree$\\
Azimuth resolution & $0.23$~m\\
Slant range resolution & $0.59$~m\\
Azimuth scene extent & $3.1$~km\\
Ground Range scene extent & $5.5$~km\\
Range bandwidth & $300$~MHz\\
Antenna bandwidth & $2589$~Hz\\
Focused azimuth bandwidth & $38275$~Hz\\
Acquisition pulse repetition frequency (PRF) & $4448$~Hz\\
Focused PRF & $42300$~Hz\\
Number of azimuth beams & $113$\\
Squint angle range & $\pm 2.2\degree$\\
Aperture time $t_{\text{AP}}$ & $7.24$~s\\
Raw data scene duration $\Delta t_{\text{raw}}$ & $7.24$~s\\
Focused scene duration $\Delta t_{\text{image}}$ & $0.43$~s\\
FM rate at scene center & $-5301$~Hz/s\\
Beam sweep rate at scene center & $-5301$~Hz/s\\
\hline
\end{tabular}
\end{table}

Due to the longer integration time of approximately $7$~s in the TerraSAR-X staring spotlight mode, several challenges arise in SAR processing \cite{pra:14}, e.g.,
\begin{enumerate*}[label=\arabic*)]
\item the stop-and-go approximation becomes invalid, i.e., satellite movement between transmitting and receiving the chirp signal can no longer be neglected;
\item satellite trajectory deviates too much from a linear track, i.e., orbit curvature needs to be taken into account;
\item tropospheric delay could vary significantly within the large squint angle span and therefore needs to be corrected.
\end{enumerate*}
All of these effects are considerately accounted for in a revised version of the TerraSAR-X multimode SAR processor \cite{bre:10, duq:15}.

InSAR processing, on the other hand, requires merely few adaptations. As in the sliding spotlight mode, the master and slave images are coregistered (resampled) on the basis of point-like scatterers in order to generate a coherent interferogram \cite{ein:09}. A requirement is the knowledge of the Doppler centroid frequency $f_{\text{DC}}$ as a function of the focused image time $t_{\text{image}}$. Since $f_{\text{DC}}$ is annotated as a (first-order) polynomial of the raw data time $t_{\text{raw}}$ in the TerraSAR-X products, it is suggested in \cite{ein:09, fri:07} to perform time conversion for the sliding spotlight datasets via
\begin{equation}
t_{\text{image}} = t_{\text{raw}} - \frac{f_{\text{DC}}(t_{\text{raw}})}{\text{FM}}.
\end{equation}
This relation, however, does not hold for the staring spotlight mode, in which the FM rate equals the beam sweep rate, i.e., a target is visible throughout the whole raw data duration. In order to circumvent this problem, a $3$-by-$3$ grid containing $f_{\text{DC}}$ in $t_{\text{image}}$ is provided as a TerraSAR-X annotation component \cite{fri:07}. Its structure is described in the appendix of this paper. This grid could be interpolated in order to derive the $f_{\text{DC}}$ at every point of the focused image, which allows considering second-order variations of $f_{\text{DC}}$ along range.

\begin{figure}[!tp]
\centering
\includegraphics[width=0.5\textwidth]{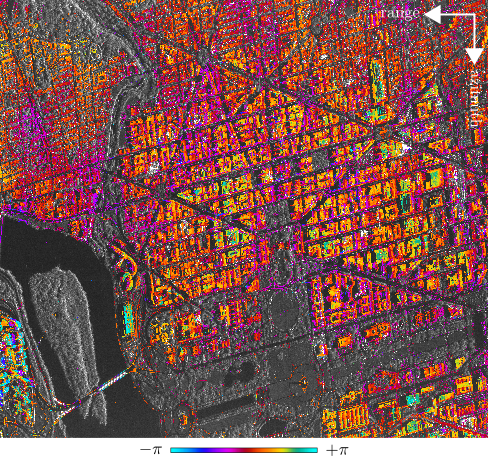}
\caption{Staring spotlight differential interferogram of Washington, D.C.\ with a spatial perpendicular baseline of approximately $-71$~m and a temporal baseline of $-22$~days.} \label{fig:3}
\end{figure}

\begin{figure}[!tp]
\centering
\includegraphics[width=0.5\textwidth]{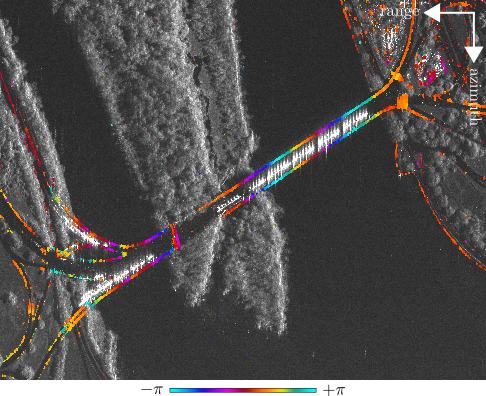}
\caption{Zoomed-in view of Fig.~\ref{fig:3} on the Theodore Roosevelt Bridge (lower left).} \label{fig:4}
\end{figure}

As an example, Fig.~\ref{fig:3} shows a differential interferogram of Washington, D.C.\ with an effective baseline of approximately $-71$~m. The master and slave scenes were acquired respectively on October 31, 2015 and October 9, 2015 and processed with the integrated wide area processor (IWAP) \cite{ada:13, rod:13}. A low-pass filtered digital elevation model (DEM) with a spatial resolution of $1$~arcsecond from the Shuttle Radar Topography Mission was used. The differential phase consists primarily of topographic phase which is related to residual height. As can be observed in Fig.~\ref{fig:4}, the Theodore Roosevelt Bridge in the lower left corner of Fig.~\ref{fig:3} is subject to spatially correlated motion, presumably due to thermal dilation and contraction between piers caused by periodical temperature change.

The next section briefly revisits the principles of TomoSAR and elucidates the processing chain which was employed to produce the results in section~\ref{sec:4} and \ref{sec:5}.

\section{TomoSAR Principles} \label{sec:3}

\begin{figure}[!tp]
\centering
\includegraphics[width=0.35\textwidth]{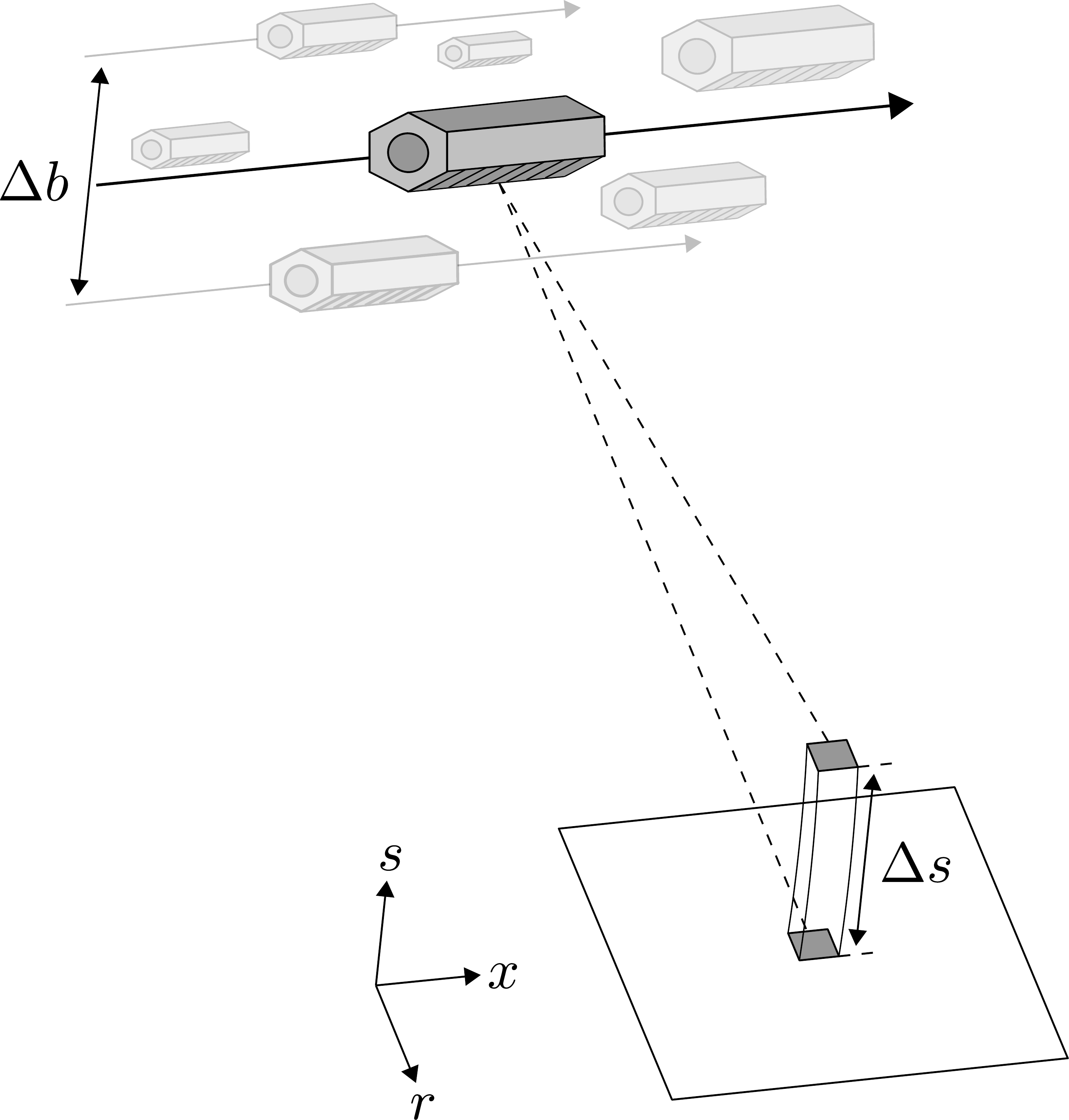}
\caption{Layover phenomenon in side-looking SAR imaging. $x$, $r$ and $s$ represent respectively azimuth, range, and elevation axes that form a local 3-D Cartesian coordinate system. An elevation aperture $\Delta b$ is built by means of repeat-pass measurements to resolve multiple scatterers in the far-field toroid segment with elevation extent $\Delta s$.} \label{fig:5}
\end{figure}

Due to the common side-looking geometry of spaceborne SAR sensors, echoes of the chirp signal from equidistant targets within an elevation extent $\Delta s$ in the far field sum to give one measurement for each azimuth-range pixel in the focused image, as illustrated in Fig.~{\ref{fig:5}}. The 3-D azimuth-range-elevation ($x$-$r$-$s$) reflectivity profile is thus embedded in 2-D, i.e., information regarding elevation is encoded during imaging. TomoSAR is a technique to reconstruct the elevation axis from multibaseline measurements \cite{rei:00, gin:02, for:05}. For spaceborne SAR, this multibaseline configuration is usually achieved by repeat-pass measurements (depicted as semitransparent satellite models in Fig.~{\ref{fig:5}}), in which scatterers' motion in the course of time often needs to be taken into account. A well-established theory models the complex InSAR measurement $g_n$ of a specific pixel in the $n$-th interferogram as the integration of a phase-modulated elevation-dependent complex reflectivity profile $\gamma(s)$ over $\Delta s$ \cite{lom:05, for:09, zhu:11}:
\begin{equation} \label{eq:2}
g_n \approx \int_{\Delta s} \gamma(s) \exp\big( -i\,2\mathrm{\pi} (\xi_n s + 2d(s, t_n)/\lambda) \big) \mathrm ds,
\end{equation}
where $\xi_n \coloneqq 2b_n/(\lambda r)$ is the elevation frequency that is proportional to the effective baseline $b_n$ ($\lambda$ and $r$ are respectively the radar wavelength and the range between sensor and target in the master image), and $d(s, t_n)$ is the line-of-sight displacement of the scatterer at elevation position $s$ and temporal baseline $t_n$. In order to reduce the number of unknowns, \(d(s, t_n)\) could be modeled as a linear combination of basis functions.
It can be shown that (\ref{eq:2}) is equivalent to a multidimensional spectral estimation problem \cite{zhu:11}. After discretizing $s$ and displacement parameters, and subsequently replacing integration by finite sum, a linear model for all $N$ InSAR measurements can be formulated as
\begin{equation} \label{eq:3}
\mathbf{g} \approx \mathbf{R} \boldsymbol\gamma,
\end{equation}
where $\mathbf{g} \coloneqq (g_1, \ldots, g_N) \in \mathbb C^N$ is the complex InSAR measurement vector, $\mathbf{R} \in \mathbb C^{N \times L}$ is the TomoSAR dictionary, and $\boldsymbol\gamma \in \mathbb C^L$ is the discrete elevation-motion reflectivity profile (or spectrum).

Various algorithms were proposed to estimate $\boldsymbol\gamma$ with given $\mathbf{R}$ and $\mathbf{g}$. A common approach is to use Tikhonov regularization \cite{zhu:10a}
\begin{equation} \label{eq:4}
\minimize_{\boldsymbol\gamma} \|\mathbf{R}\boldsymbol\gamma - \mathbf{g}\|_2^2 + \delta \|\boldsymbol\gamma\|_2^2,
\end{equation}
where $\delta>0$ is a regularization constant. Note that (\ref{eq:4}) is equivalent to the maximum a posteriori estimator of $\boldsymbol\gamma$ provided that the measurement noise is additive and white with variance $\delta$, and $\boldsymbol\gamma$ is white with variance $1$.

If one is primarily concerned with man-made objects in high-resolution spotlight images acquired over urban areas, it is deemed reasonable to assume that radar echoes in the far field are dominated by those from merely few point-like scatterers within the toroid segment in Fig.~\ref{fig:5}, i.e., $\boldsymbol\gamma$ is presumed to be compressible and thus $\mathbf{g}$ could be sufficiently approximated by a linear combination of few atoms (columns) of $\mathbf{R}$. This hypothesis gave rise to approaches with sparsity-driven $\ell_1$ regularization \cite{zhu:10b, bud:11}:
\begin{equation} \label{eq:5}
\minimize_{\boldsymbol\gamma} \|\mathbf{R}\boldsymbol\gamma - \mathbf{g}\|_2^2 + \epsilon \|\boldsymbol\gamma\|_1,
\end{equation}
where $\epsilon>0$ is another regularization constant.

In terms of the capability to resolve multiple point-like scatterers, conventional methods such as Tikhonov regularization (\ref{eq:4}) are limited by the elevation resolution $\rho_s \coloneqq \lambda r/(2\Delta b)$, where $\Delta b$ is the elevation aperture as shown in Fig.~\ref{fig:5}. For TerraSAR-X, $\rho_s$ is in the order of several tens of meters (typically \(20\)--\(30\) m given a sufficiently large stack), as a consequence of the satellite being confined to a $250$-m orbit tube \cite{yoo:09}. Given one single scatterer within the resolution cell, a lower bound on the errors of elevation estimates $\hat s$ can be derived as \cite{bam:09}
\begin{equation} \label{eq:6}
\sigma_{\hat s} \coloneqq \frac{\lambda r}{4\pi \sqrt{N} \sqrt{2\mathit{SNR}} \; \sigma_b},
\end{equation}
where $\mathit{SNR}$ is the scatterer's signal-to-noise ratio, and $\sigma_b$ is the standard deviation of effective baselines. In case of double scatterers, their mutual interference could be modeled as a scaling factor which depends primarily on their elevation distance and phase difference \cite{zhu:12}. For TerraSAR-X, this lower bound is approximately one order smaller than $\rho_s$ and could be approached by means of $\ell_1$ regularization (\ref{eq:5}). In other words, (\ref{eq:5}) could achieve superresolution \cite{zhu:14}.

\begin{figure}[!tp]
\centering
\includegraphics[width=0.5\textwidth]{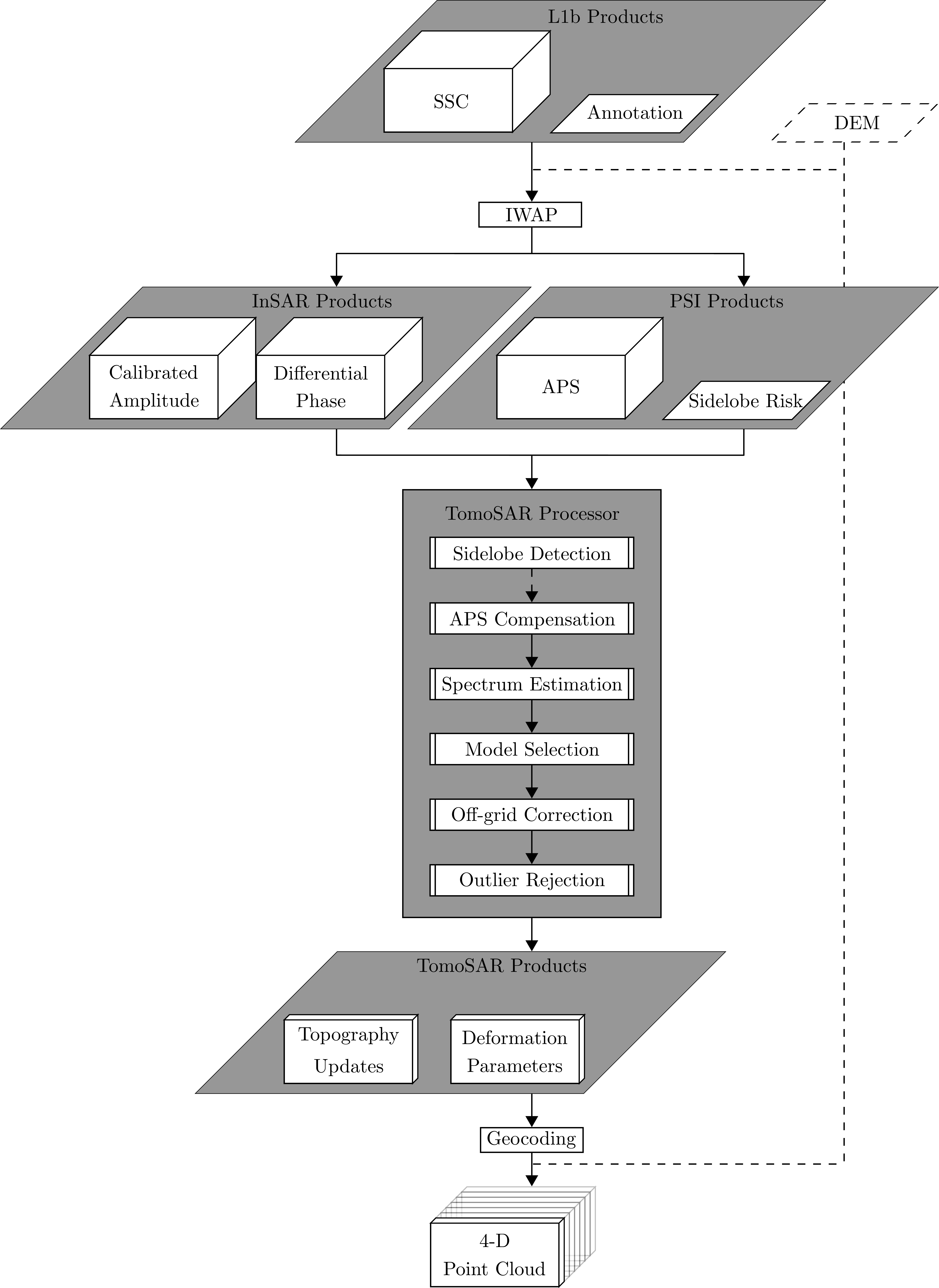}
\caption{Top-down model of the processing chain. Modified from \cite{wan:14}.}
\label{fig:6}
\end{figure}

As an overview, a top-down model of the processing chain is illustrated in Fig.~\ref{fig:6} and consists primarily of the following parts:
\begin{enumerate}
\item \emph{Preprocessing} (via IWAP), which takes focused single-look slant-range complex (SSC) images as input and performs
\begin{enumerate}
\item \emph{InSAR processing}, which provides raster images of calibrated amplitude and differential phase, and subsequently
\item \emph{PSI processing}\label{list:1b}, which estimates atmospheric phase screen (APS) from single point-like targets and a sidelobe risk map \cite{ada:09,ada:13,igl:13}.
\end{enumerate}
Note that the use of a DEM is optional if the concerned terrain is relatively flat.

\item \emph{TomoSAR processing.}
\begin{enumerate}
\item \emph{Sidelobe detection}. A simple hypothesis test (thresholding) is applied to the sidelobe risk map from \ref{list:1b}).
\item \emph{APS compensation}. The estimated APS is compensated in differential phase, if the corresponding pixel concerned is, with high probability, not dominated by a sidelobe.
\item \emph{Spectrum estimation}\label{list:2c}. The elevation-motion spectrum is estimated with, for example, (\ref{eq:4}) or (\ref{eq:5}).
\item \emph{Model selection}. By minimizing the penalized negative log-likelihood, the number of scatterers is estimated to reduce false positive rate \cite{zhu:12}. If $\ell_1$ regularization is employed in \ref{list:2c}), the underestimated amplitude is hereby corrected as a byproduct.
\item \emph{Off-grid correction}. In order to ameliorate the off-grid problem as a consequence of discretizing elevation and motion parameters, the estimated elevation-motion spectrum from \ref{list:2c}) is oversampled in a neighborhood of each statistically significant scatterer. A local maximum is detected in the oversampled high-dimensional signal, which allows better quantization.
\item \emph{Outlier rejection}. As a natural extension of the complex ensemble coherence for single point-like scatterers \cite{fer:01}, we define for the multiple-scatterer case
\begin{equation} \label{eq:7}
\eta \coloneqq \frac{1}{N} \sum_{n=1}^N \exp\big(-i\,(\angle \mathbf{r}^n\boldsymbol\gamma - \angle g_n) \big),
\end{equation}
where $\angle:\mathbb{C} \to \mathbb{R}$ returns the phase of a complex number, and $\mathbf{r}^n$ denotes the $n$-th row of the TomoSAR dictionary $\mathbf{R}$. We reject outliers, i.e., scatterers whose phase history deviates significantly from the adopted model, by thresholding of $|\eta|$.
\end{enumerate}

\item \emph{Postprocessing}, which couples the updated topography and its deformation parameters to produce a 4-D geocoded point cloud.
\end{enumerate}

In the next section, we demonstrate for the first time TerraSAR-X staring spotlight TomoSAR results produced with the abovementioned processing chain. Based on a sufficient number of acquisitions, the demonstration is given not only for individual urban infrastructures, but also on the scale of a city.

\section{First Practical Demonstration of Staring Spotlight TomoSAR} \label{sec:4}

\begin{figure}[!tp]
\centering
\includegraphics[width=0.5\textwidth]{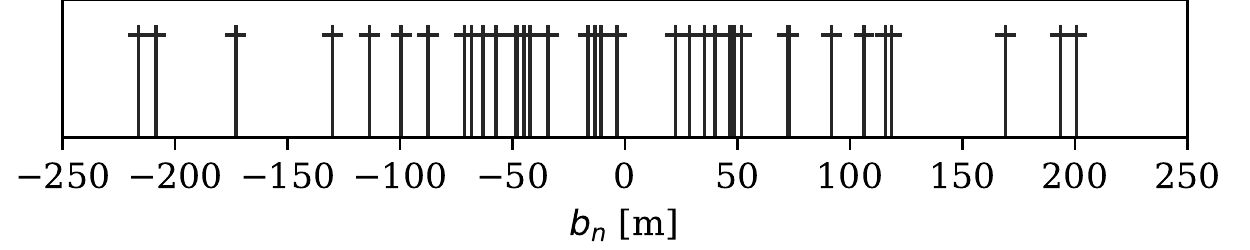}
\caption{Distribution of effective baselines $b_n$.}
\label{fig:7}
\end{figure}

Forty-one staring spotlight images were acquired by TerraSAR-X from July 4, 2014 to November 30, 2016 with a constant repeat interval of $22$ days, i.e., every second orbit. The image from October 31, 2015 with an incidence angle of $40.7\degree$ at scene center was chosen as the master due to its central position in the spatial-temporal baseline plot and relatively small atmospheric delays. Fig.~\ref{fig:7} shows the distribution of effective baselines $b_n$ with respect to the master scene, which are indeed confined to $\pm 250$~m. The elevation aperture $\Delta b$ is approximately $417$~m, which leads to an elevation resolution $\rho_s$ of approximately $24.6$~m at scene center. Given an $\mathit{SNR}$ of $2$~dB, the lower bound for single point-like scatterers $\sigma_{\hat s}$ is merely $1.44$~m, i.e., less than $6\%$ of $\rho_s$.

As previously mentioned in section \ref{sec:3}, the preprocessing (i.e., InSAR and PSI processing) was accomplished by IWAP. In order to decrease the computational cost, we exclusively considered the pixels with SCR $\geq 1.7$~dB as candidates for TomoSAR processing, i.e., heavily vegetated areas and water bodies were likely masked out. The number of candidates was further reduced by eliminating those pixels, each of which has an estimated likelihood of being a sidelobe larger than $0.45$. As a result, we only processed approximately $12\%$ of the original raster data. Scatterers' motion was modeled with a coupled linear and sinusoidal model with the latter having a period of one year. The elevation-motion spectrum was estimated either with Tikhonov regularization (\ref{eq:4}) for the whole scene, or with $\ell_1$ regularization (\ref{eq:5}) for certain regions of interest. The maximum number of point-like scatterers within each resolution cell was set to $2$ and the model selector was trained such that the false positive rate for double scatterers, i.e., the empirical probability that two scatterers are detected whereas there is at most one, is below $1\text\textperthousand$. A neighborhood of each selected scatterer in its 3-D elevation-motion ($s$-$v$-$a$, where $v$ is the linear deformation rate and $a$ is the periodical deformation amplitude) spectrum was oversampled with a factor of $10$ to alleviate the off-grid problem. Scatterers with an ensemble coherence (\ref{eq:7}) lower than $0.6$ were considered as outliers and excluded from postprocessing.

\begin{figure*}[!tp]
\centering
\subfloat[Updated topography $h \text{~[m]}$]{\includegraphics[width=1.\textwidth]{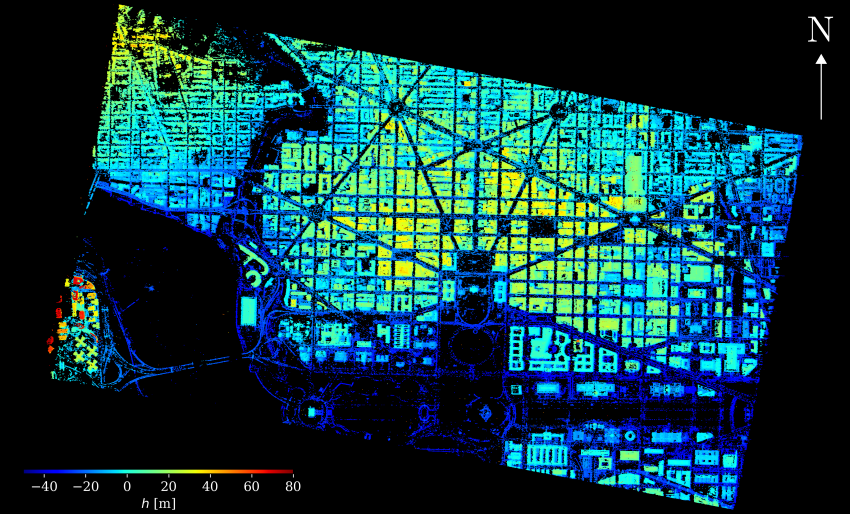}
\label{fig:8a}}
\hfil
\subfloat[Linear deformation rate $v\text{~[mm/year]}$]{\includegraphics[width=1.\textwidth]{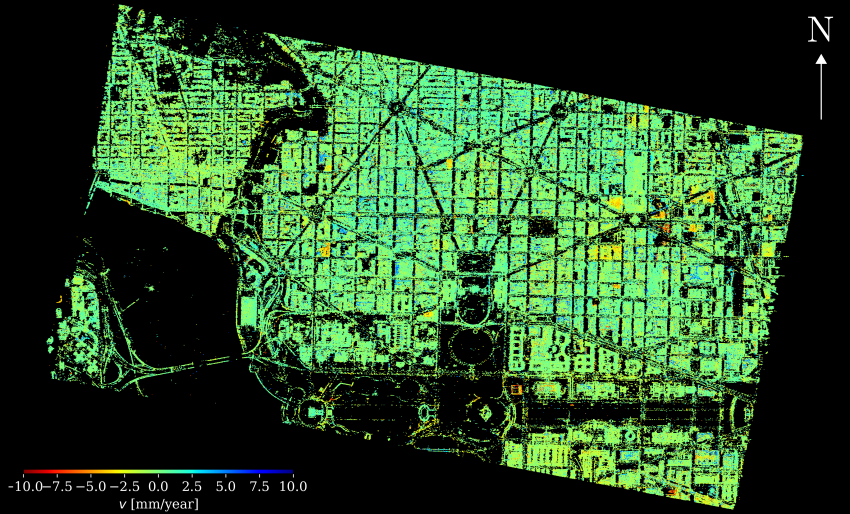}
\label{fig:8b}}
\caption{TomoSAR results of Washington, D.C. with $41$ TerraSAR-X staring spotlight acquisitions.}
\label{fig:8}
\end{figure*}

The updated topography $h$, linear deformation rate $v$ and periodical deformation amplitude $a$ are shown in Fig.~\ref{fig:8a}, \ref{fig:8b} and \ref{fig:8c}, respectively. On the Potomac River (lower left), scarcely any point-like scatterers could be detected, except for those from the National Memorial on the Theodore Roosevelt Island (cf.\ Fig.~\ref{fig:3}), and those on the Theodore Roosevelt Bridge (cf.\ Fig.~\ref{fig:4}). The National Mall in the lower part is in general void of point-like scatterers due to its vegetation.

Most of the buildings in the scene appear to be flat with the exception of several high-rise ones in Rosslyn, Virginia (lower left, to the west of the Theodore Roosevelt Bridge). Zoomed-in views of the Watergate complex and the John F.\ Kennedy Center for the Performing Arts are provided as Fig.~\ref{fig:9} and \ref{fig:10}, respectively. Due to the limitations of Google Earth merely $6\%$ of the original point cloud was used for visualization. 

Bridges and overpasses are in general subject to periodical deformation as a result of temperature changes, i.e., dilation between piers or fixed bearings in summer and contraction in winter. The estimated periodical deformation amplitude of the Theodore Roosevelt Bridge is shown in Fig.~\ref{fig:11}. As an example, Fig.~\ref{fig:12} demonstrates the phase history of two scatterers within a resolution cell. The higher scatterer (depicted as red dot) is located on the bridge, while the lower (blue) resides at one of the piers. The estimated height difference of these two scatterers is approximately $8.3$~m, which lies in the superresolution regime. As the upper right plot of Fig~\ref{fig:12} suggests, the lower scatterer on the pier undergoes little deformation, whereas the periodical deformation amplitude of the higher scatterer on the bridge was estimated to be approximately $2.9$~mm. The topography and deformation model of double scatterers fits quite well to the InSAR measurements (see the lower right plot of Fig~\ref{fig:12}) and the ensemble coherence amounts to approximately $0.97$.

The Washington Marriott Marquis hotel (opened on May 1, 2014) beside the Walter E.\ Washington Convention Center appears to suffer from subsidence that is presumably due to the building weight (see Fig.~\ref{fig:13a}). In addition, it undergoes thermal dilation and contraction which are more significant on roof than on facade, as can be observed in Fig.~\ref{fig:13b}. Fig.~\ref{fig:14} shows the resolved layover effect of two scatterers, which is a typical case of roof-facade interaction. The higher and lower scatterers subside with a linear rate of $-1.1$ and $-1.0$~mm/year, respectively. The scatterer on the roof moves periodically with an amplitude of approximately $3.0$~mm, while on the contrary the one on the facade is subject to little such deformation. Similar to the previous example in Fig.~\ref{fig:12}, the TomoSAR model could describe the phase history sufficiently well with an ensemble coherence of approximately $0.97$.

As one last example, Fig.~\ref{fig:15a} and \ref{fig:15b} show the updated topography and periodical deformation amplitude of the Rosslyn Twin Towers, respectively. Clearly the amplitude of thermal dilation and contraction is highly correlated with building height. Note that the tower on the left has smaller point density on the left-hand side of the facade due to its convex shape as seen from the radar wavefront. Fig.~\ref{fig:16} demonstrates another typical case of layover effect in urban areas which is the facade-ground (or facade-lower-infrastructure) interaction. The periodical deformation amplitude of the higher and lower scatterers were estimated to be approximately $5.0$ and $2.0$~mm, respectively.

The next section reports a preliminary comparison of sliding and staring spotlight TomoSAR using TerraSAR-X data. The comparison is based on a limited number of acquisitions and therefore restricted to two small typical urban areas.

\setcounter{figure}{7}

\begin{figure*}[!tp]
\centering
\setcounter{subfigure}{2}
\subfloat[Periodical deformation amplitude $a\text{~[mm]}$]{\includegraphics[width=1.\textwidth]{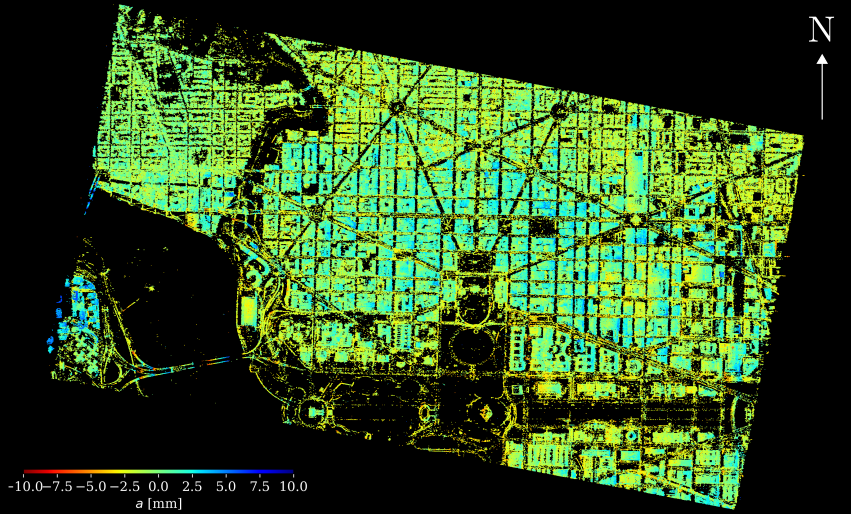}
\label{fig:8c}}
\caption{TomoSAR results of Washington, D.C. with $41$ TerraSAR-X staring spotlight acquisitions (continued).}
\end{figure*}

\begin{figure*}[!tp]
\centering
\subfloat{\includegraphics[width=1.\textwidth]{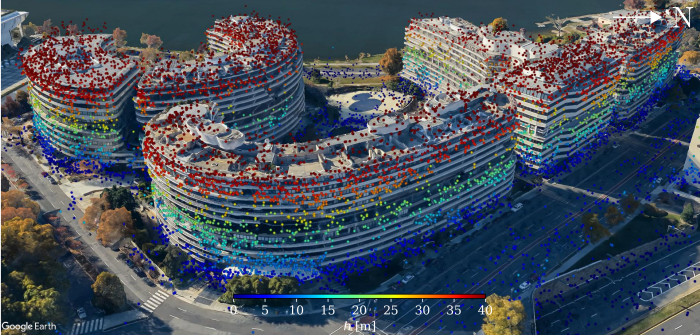}}
\caption{$6\%$ of the original point cloud of the Watergate complex that is overlaid on Google Earth 3-D photo-realistic building model and color-coded by updated topography $h$ [m].} \label{fig:9}
\end{figure*}

\begin{figure*}[!tp]
\centering
\subfloat{\includegraphics[width=1.\textwidth]{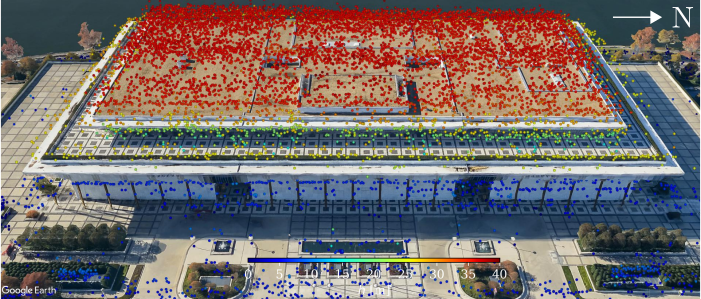}}
\caption{$6\%$ of the original point cloud of the John F.\ Kennedy Center for the Performing Arts that is overlaid on Google Earth 3-D photo-realistic building model and color-coded by updated topography $h$ [m].}
\label{fig:10}
\end{figure*}




\begin{figure*}[!tp]
\centering
\subfloat{\includegraphics[width=1.\textwidth]{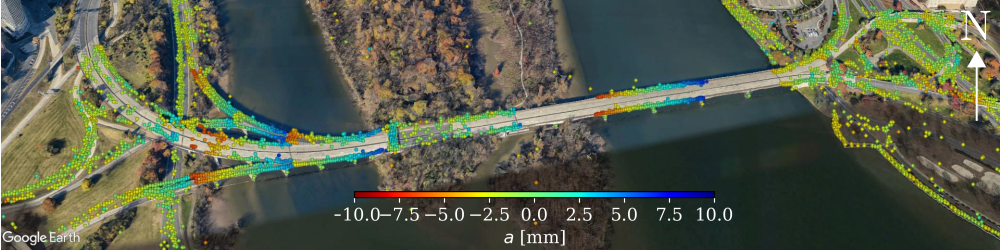}}
\caption{$6\%$ of the original point cloud of the Theodore Roosevelt Bridge that is overlaid on Google Earth 3-D photo-realistic building model and color-coded by periodical deformation amplitude $a$ [mm].}
\label{fig:11}
\end{figure*}

\begin{figure}[!tp]
\centering
\includegraphics[width=.5\textwidth]{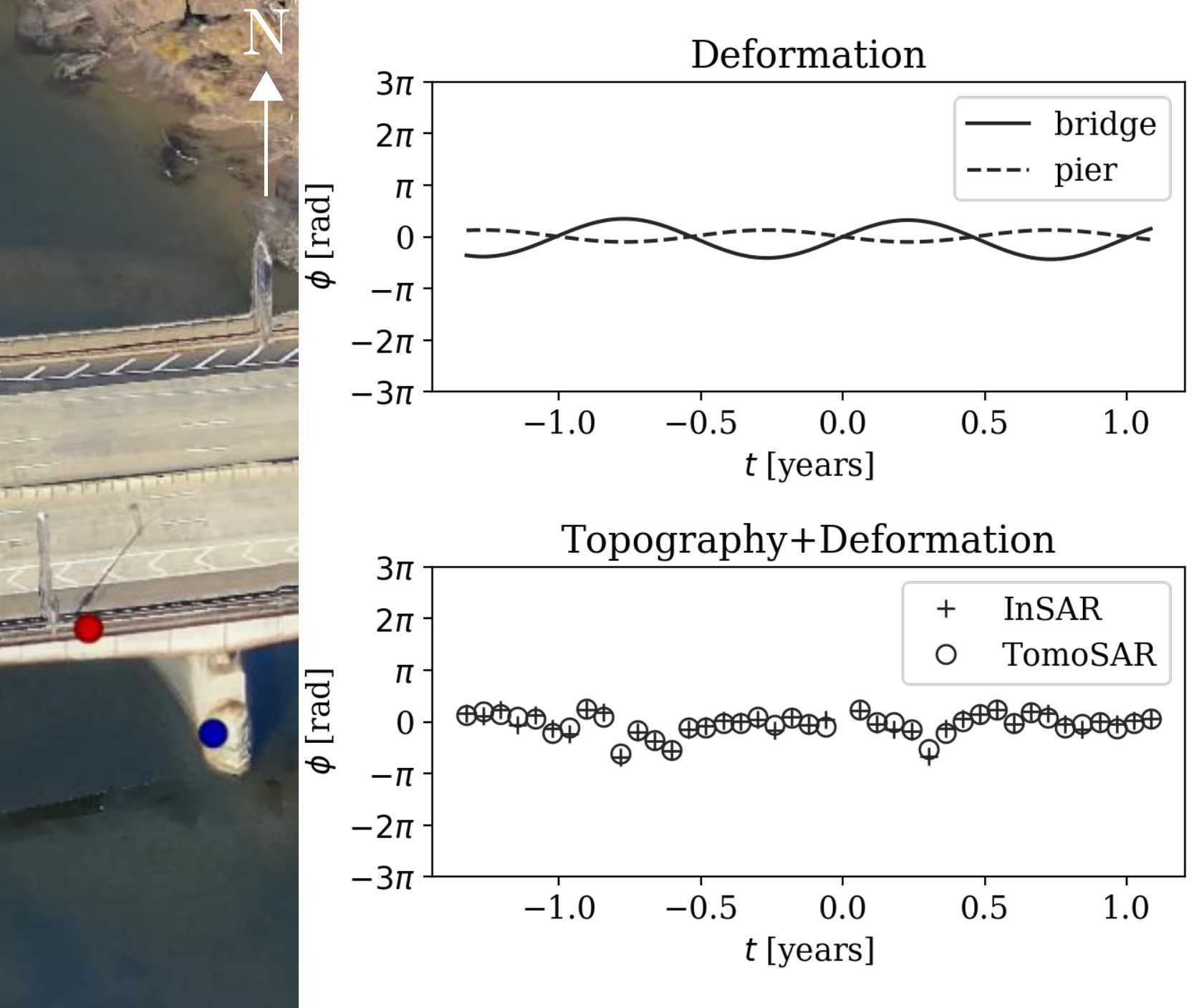}
\caption{Phase history of InSAR measurements and TomoSAR reconstruction of double scatterers subject to layover in Fig.~\ref{fig:11}. The higher and lower scatterers are marked as red and blue, respectively.}
\label{fig:12}
\end{figure}

\begin{figure}[!tp]
\centering
\subfloat[Linear deformation rate $v \text{~[mm/year]}$]{\includegraphics[width=.5\textwidth]{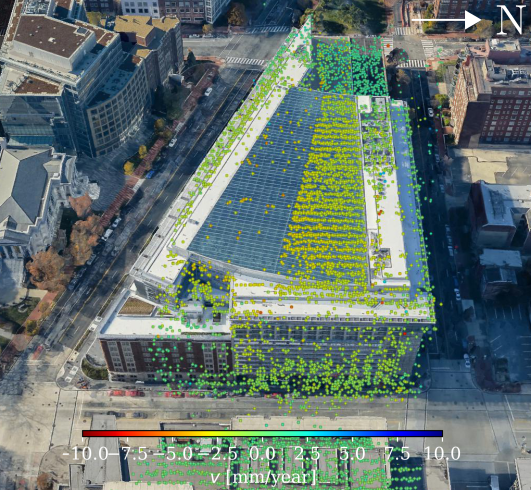}
\label{fig:13a}}
\hfil
\subfloat[Periodical deformation amplitude $a \text{~[mm]}$]{\includegraphics[width=.5\textwidth]{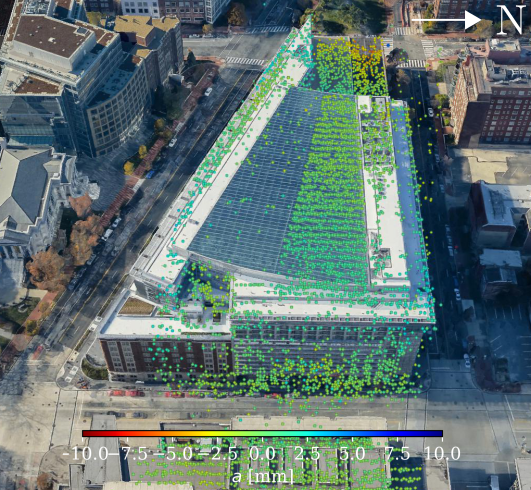}
\label{fig:13b}}
\caption{$4\%$ of the original point cloud of the Washington Marriott Marquis hotel that is overlaid on Google Earth 3-D photo-realistic building model.}
\label{fig:13}
\end{figure}

\begin{figure}[!tp]
\centering
\includegraphics[width=.5\textwidth]{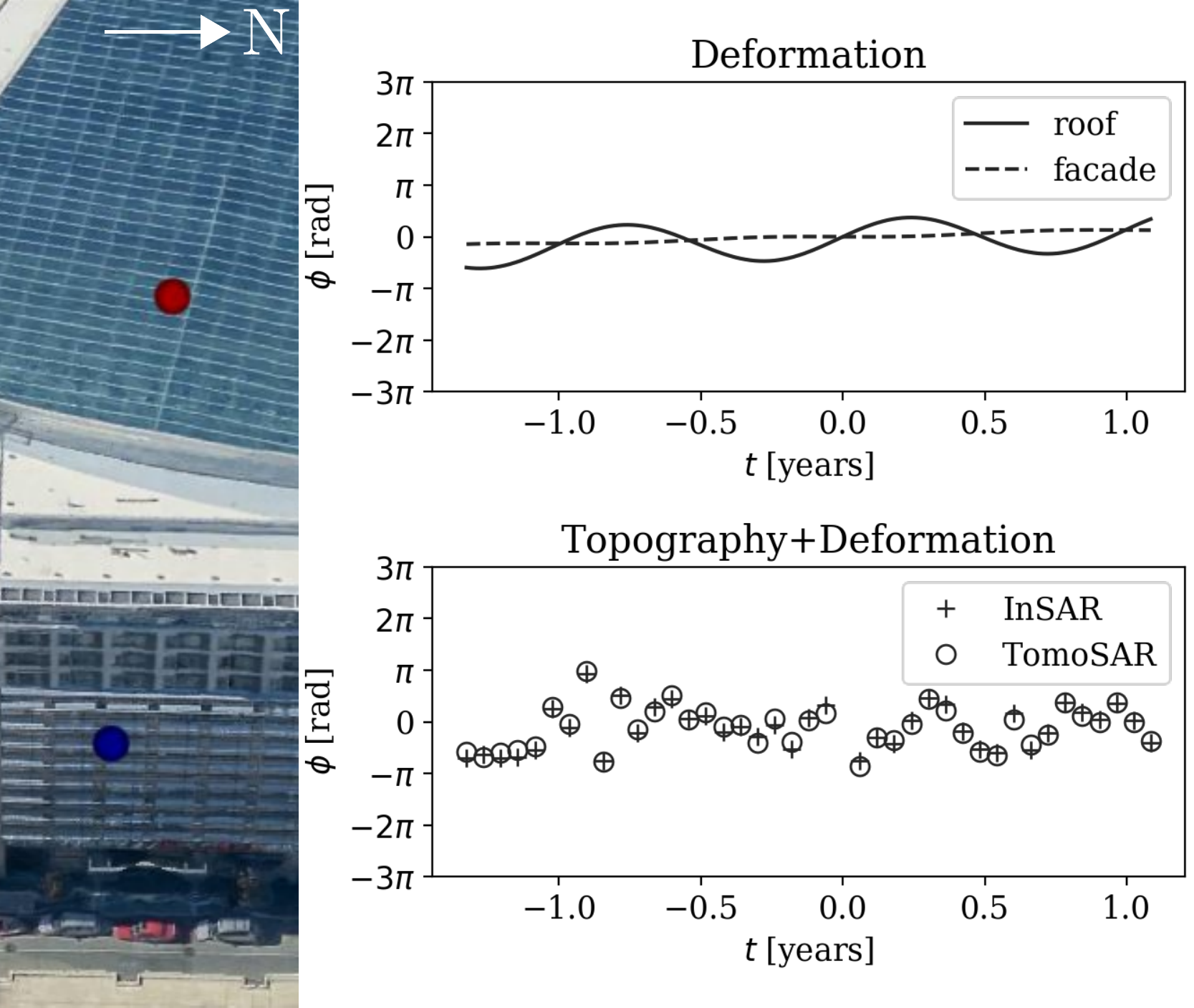}
\caption{Phase history of InSAR measurements and TomoSAR reconstruction of double scatterers subject to layover in Fig.~\ref{fig:13}. The higher and lower scatterers are marked as red and blue, respectively.}
\label{fig:14}
\end{figure}

\begin{figure}[!tp]
\centering
\subfloat[Updated topography $h \text{~[m]}$]{\includegraphics[width=.48\textwidth]{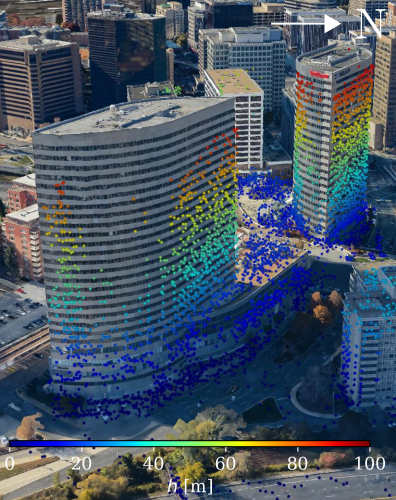}
\label{fig:15a}}
\hfill
\subfloat[Periodical deformation amplitude $a \text{~[mm]}$]{\includegraphics[width=.48\textwidth]{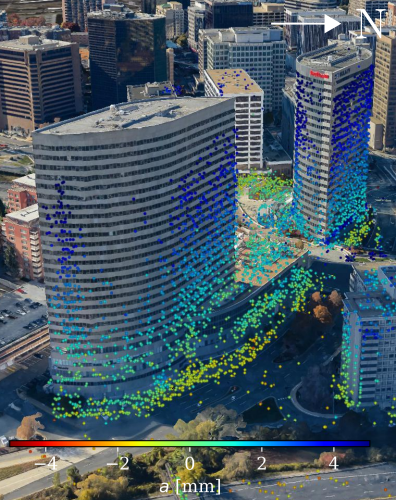}
\label{fig:15b}}
\caption{$5\%$ of the original point cloud of the Rosslyn Twin Towers that is overlaid on Google Earth 3-D photo-realistic building model.}
\label{fig:15}
\end{figure}

\begin{figure}[!tp]
\centering
\includegraphics[width=.5\textwidth]{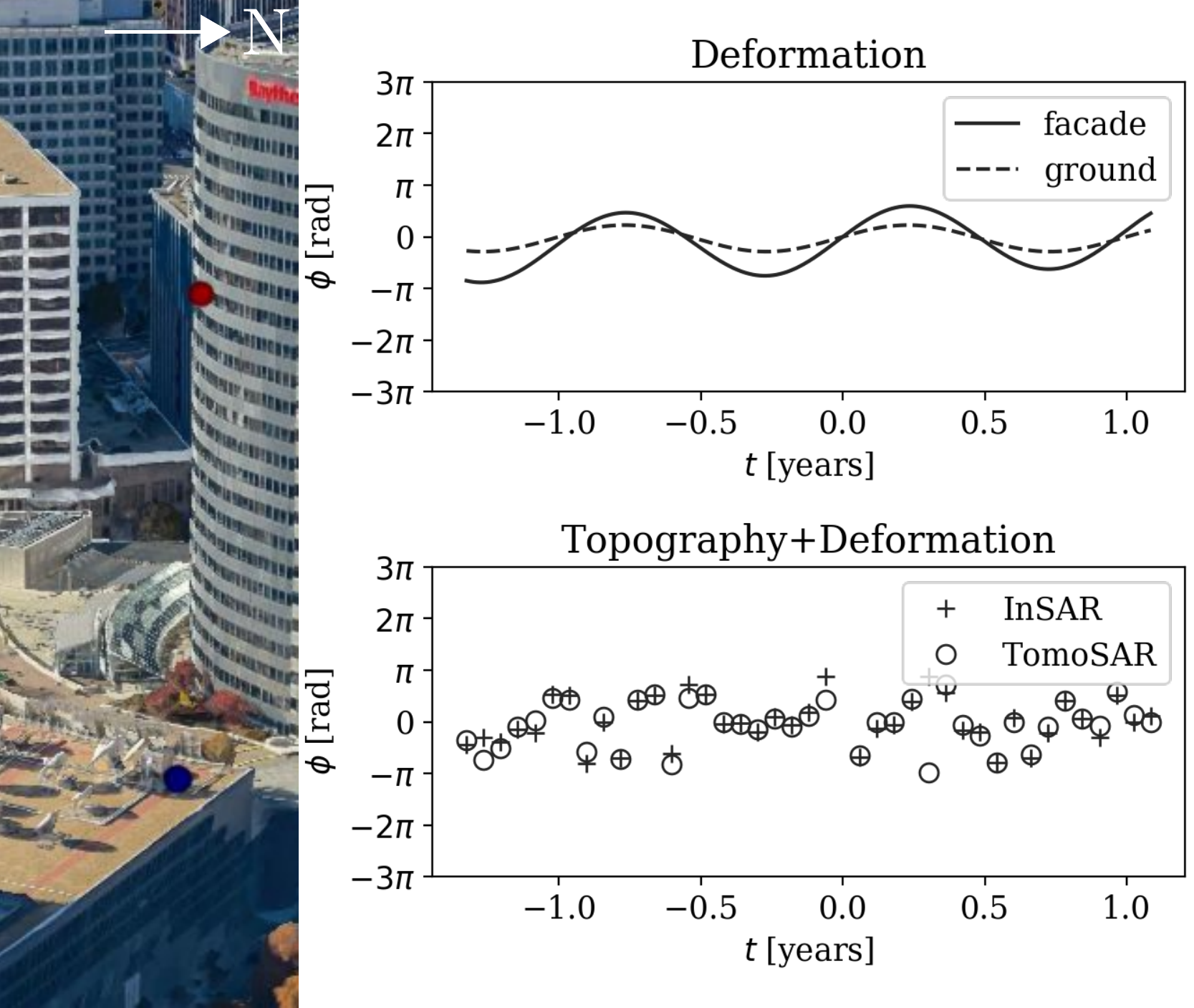}
\caption{Phase history of InSAR measurements and TomoSAR reconstruction of double scatterers subject to layover in Fig.~\ref{fig:15}. The higher and lower scatterers are marked as red and blue, respectively.}
\label{fig:16}
\end{figure}

\section{Preliminary Comparison of Sliding and Staring Spotlight TomoSAR} \label{sec:5}

Due to data unavailability, a direct comparative study of both modes was not possible for Washington D.C. Instead, we drew the comparison with two small descending interferometric stacks of the City of Las Vegas. Each stack contains \(12\) images which were acquired alternately from October, 2014 to February, 2015 during the TanDEM-X Science Phase \cite{haj:14}. For each mode, \(11\) interferograms were generated with a similar baseline distribution as in Fig.~\ref{fig:7}.

\begin{figure}[!tp]
\centering
\subfloat[Sliding]{\includegraphics[width=.5\textwidth]{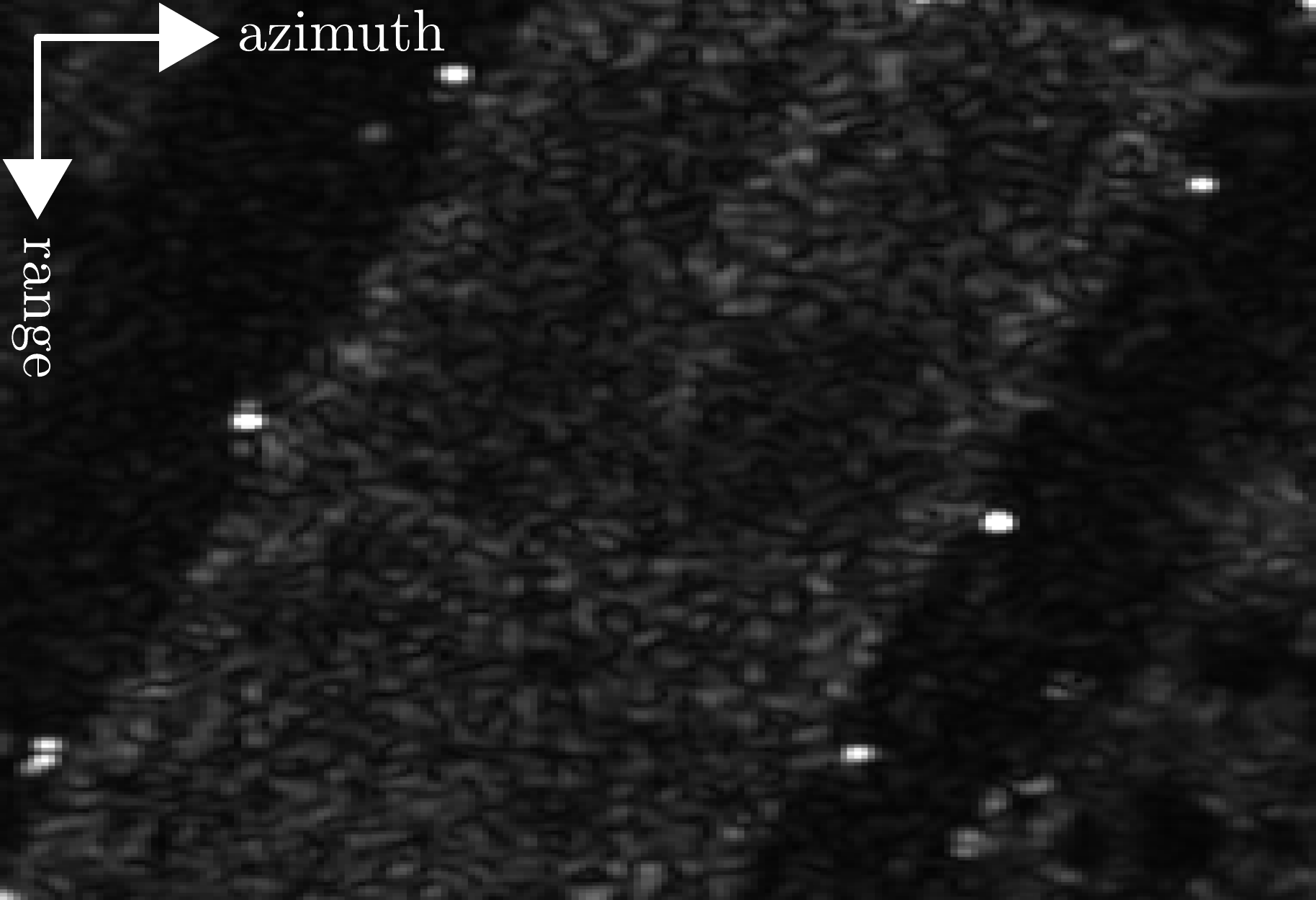}
\label{fig:17a}}
\hfil
\subfloat[Staring]{\includegraphics[width=.5\textwidth]{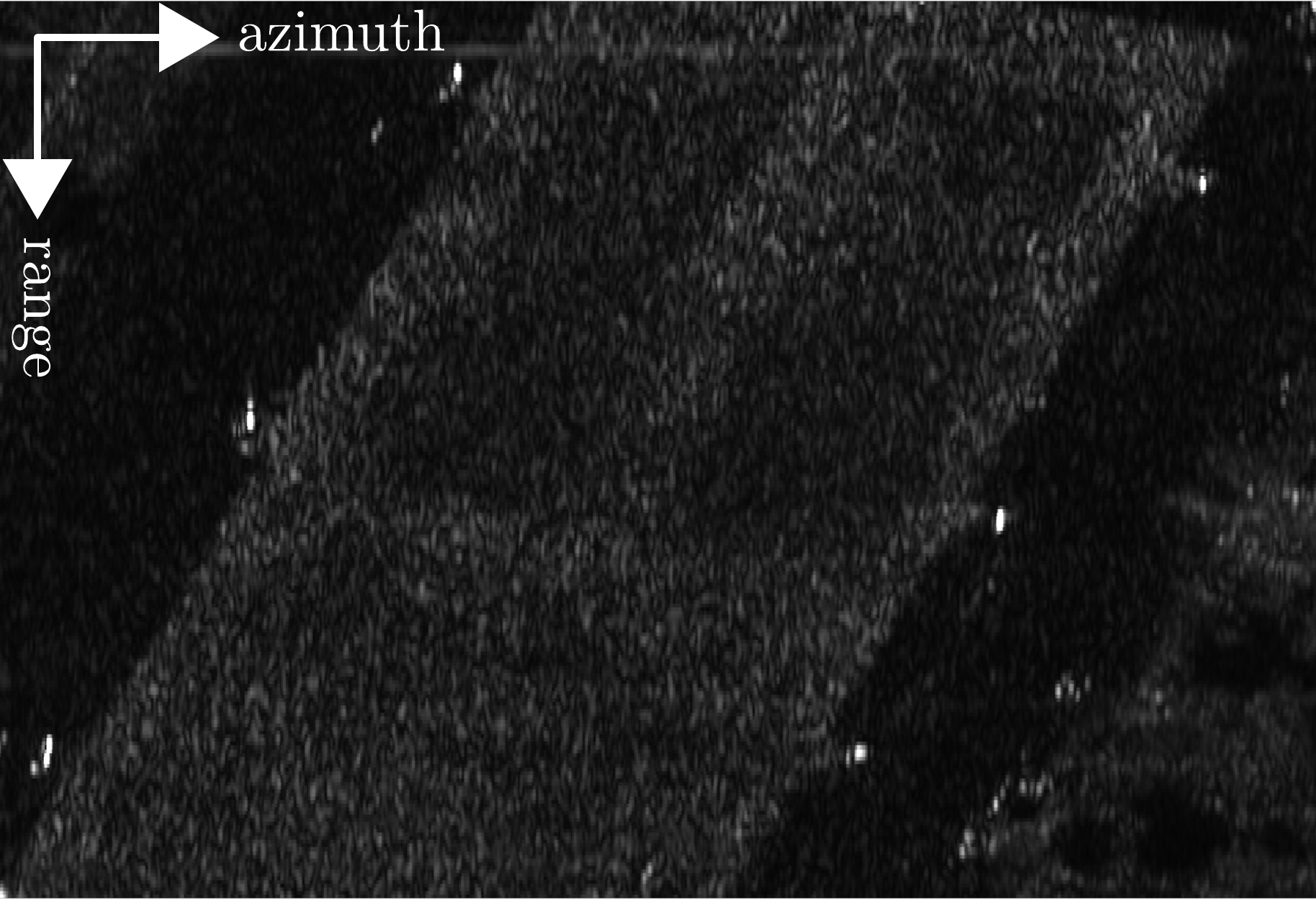}
\label{fig:17b}}
\caption{Mean intensity map of a relatively flat area in the (a) sliding and (b) staring spotlight modes.}
\label{fig:17}
\end{figure}

\begin{figure}[!tp]
\centering
\subfloat[Sliding]{\includegraphics[width=.5\textwidth]{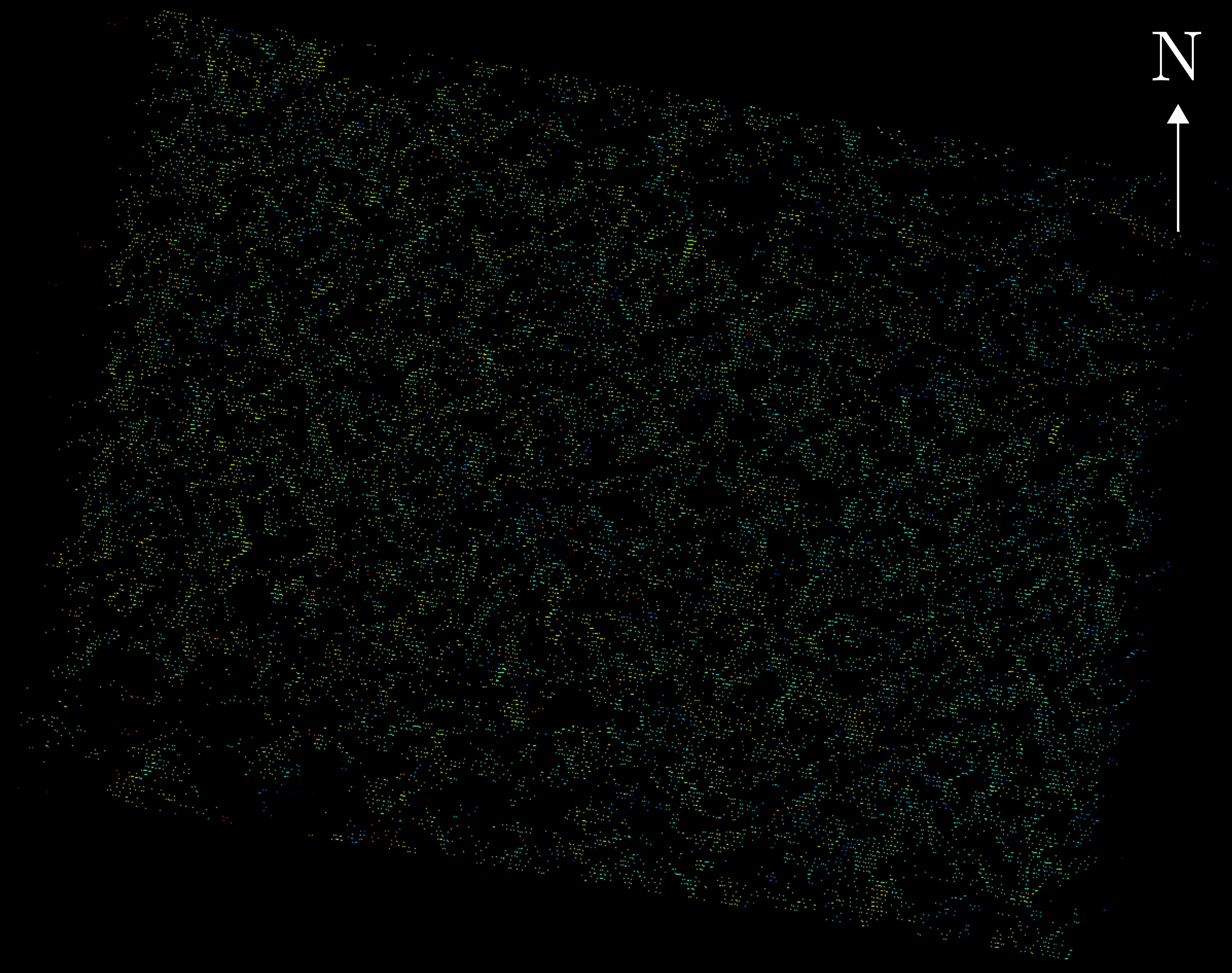}
\label{fig:18a}}
\hfil
\subfloat[Staring]{\includegraphics[width=.5\textwidth]{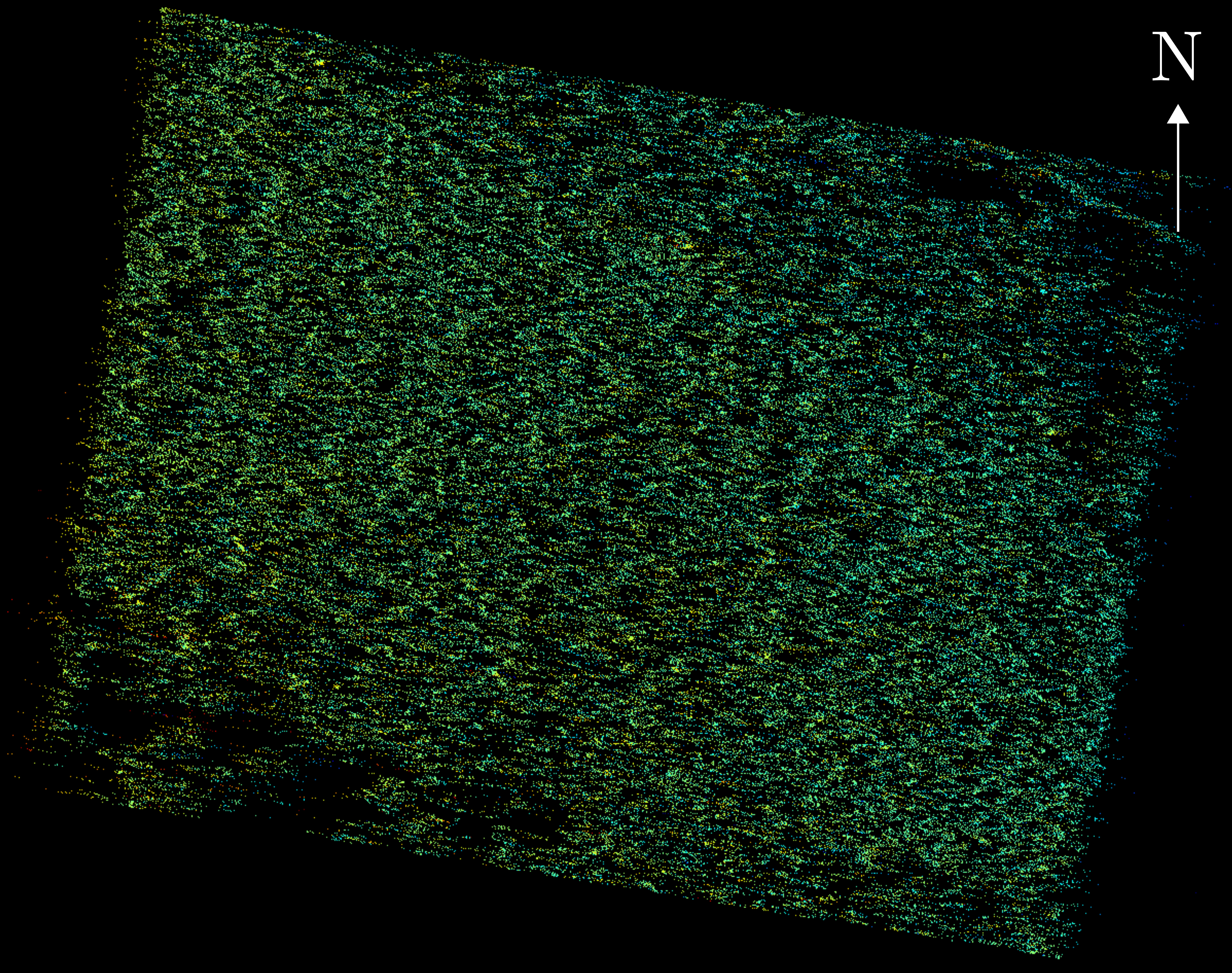}
\label{fig:18b}}
\hfil
\subfloat{\includegraphics[width=.3\textwidth]{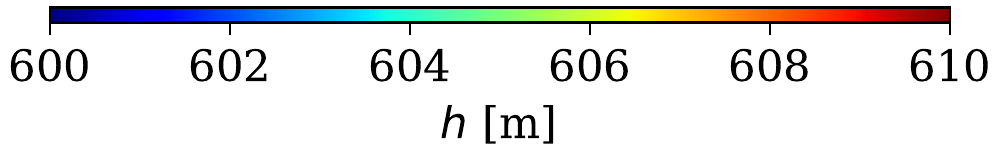}}
\caption{Updated topography \(h\) [m] of the area in Fig.~\ref{fig:17} with \(12\) TerraSAR-X images in the (a) sliding and (b) staring spotlight modes, respectively.}
\label{fig:18}
\end{figure}

Two small areas were selected for the comparison of sliding and staring spotlight TomoSAR. One of them is a relatively flat area of approximately \(0.01\) km\(^2\). The same area of interest was cropped in both datasets using ground control points. Fig.~\ref{fig:17} shows the mean intensity map in each mode. In the staring spotlight case, point-like targets appear more focused, which indicates an increase of SCR. As a result, the contrast between areas of different degrees of smoothness becomes larger, i.e., the boundaries of the rectangular surfaces in the middle of the image are much easier to recognize. The reconstructed TomoSAR point cloud is shown in Fig.~\ref{fig:18}. An increase in the number of points in the staring spotlight mode is obvious. Indeed, the point density in the staring spotlight case is approximately \(5.5\) times as high, see Tab.~\ref{tab:2}.

The assessment of the relative height accuracy is explained as follows. Since this area is relatively flat (as confirmed by Fig.~\ref{fig:18}), we fitted a plane with robust measure through each point cloud and considered it as partial ground truth. Note that this also took the local slope into account. Subsequently, we calculated the distance of each scatterer to the fitted plane and projected it into the vertical direction. In this context, we refer to the median absolute deviation of height estimate errors relative to this fitted plane as relative height accuracy.

Let us denote the vectors containing the geographic coordinates of all \(m\) scatterers as \(\bftx, \bfty, \bftz \in \bbR^m\), respectively. We seek a plane parametrized by \(\tilde a, \tilde b, \tilde c, \tilde d \in \bbR\) such that,
\begin{equation} \label{eq:8}
\tilde a \tilde x + \tilde b \tilde y + \tilde c \tilde z + \tilde d \approx 0,
\end{equation}
for each scatterer at the coordinates \(\tilde x \in \bftx\), \(\tilde y \in \bfty\), \(\tilde z \in \bftz\). Without loss of generality, let us assume that \(\tilde c = 1\). The plane fitting problem can be formulated as
\begin{equation} \label{eq:9}
\minimize_{\bfx} \|\bfA \bfx - \bfb\|_1,
\end{equation}
where \(\bfA \coloneqq
\begin{pmatrix}
\bftx & \bfty & \bf1
\end{pmatrix}
\in \bbR^{m \times 3}\), \(\bf1\) is an \(m\)-dimensional vector of ones, \(\bfx \coloneqq
\begin{pmatrix}
\tilde a & \tilde b & \tilde d
\end{pmatrix}^{\T}
\in \bbR^3\), and \(\bfb \coloneqq -\bftz\). The \(\ell_1\) loss function is known for its robustness against outliers \cite{boy:04}. Let \(\bfx^*\) denote an optimal solution and \(\mathbf n \coloneqq
\begin{pmatrix}
x_1^* & x_2^* & 1
\end{pmatrix}^{\T}
\) be a corresponding plane normal, the signed distance of scatterers to the fitted plane is given by \((\bfA \bfx^* + \bftz) / \|\bfn\|_2\). Due to the large scale of problem (\ref{eq:9}), i.e., \(m > 10^5\) as shown in Tab.~\ref{tab:2}, generic conic solvers may not be able to solve it efficiently. Based on the alternating direction method of multipliers (ADMM) \cite{boy:11}, we developed a fast solver with super-linear convergence rate, see Algorithm \ref{alg:1}, where \(\bfz\) and \(\bfy\) are respectively auxiliary primal and dual variables, \(\rho>0\) is a penalty parameter for a smoothness term in the augmented Lagrangian (fixed to \(1\) in this paper), and \(\prox_{\ell_1,\lambda}(\bfw) \coloneqq (\bfw - \lambda)_+ - (-\bfw - \lambda)_+\) is the elementwise soft thresholding operator \cite{par:14}, where \((\bfu)_+ \coloneqq \max(\bfu,0)\) replaces the negative entries with zeros.

\begin{algorithm}[H]
\caption{ADMM-based algorithm for solving (\ref{eq:9})} \label{alg:1}
\begin{algorithmic}[1]
\State \textbf{Input:} \(\bfA\), \(\bfb\), \(\rho\)
\State \textbf{Initialize} \Let{\bfz}{\bf0}, \Let{\bfy}{\bf0}
\State \textbf{Until} stopping criterion is satisfied, \textbf{Do}
\State \quad \Let{\bfx}{ (\bfA^{\T} \bfA)^{-1} \big(\bfA^{\T} (\bfb + \bfz - \frac{1}{\rho}\bfy)\big) }
\State \quad \Let{\bfz}{\prox_{\ell_1, 1/\rho}(\bfA\bfx - \bfb + \frac{1}{\rho}\bfy)}
\State \quad \Let{\bfy}{\bfy + \bfA\bfx - \bfb - \bfz}
\State \textbf{Output:} \(\bfx\)
\end{algorithmic}
\end{algorithm}

\begin{figure}[!tp]
\centering
\includegraphics[width=.5\textwidth]{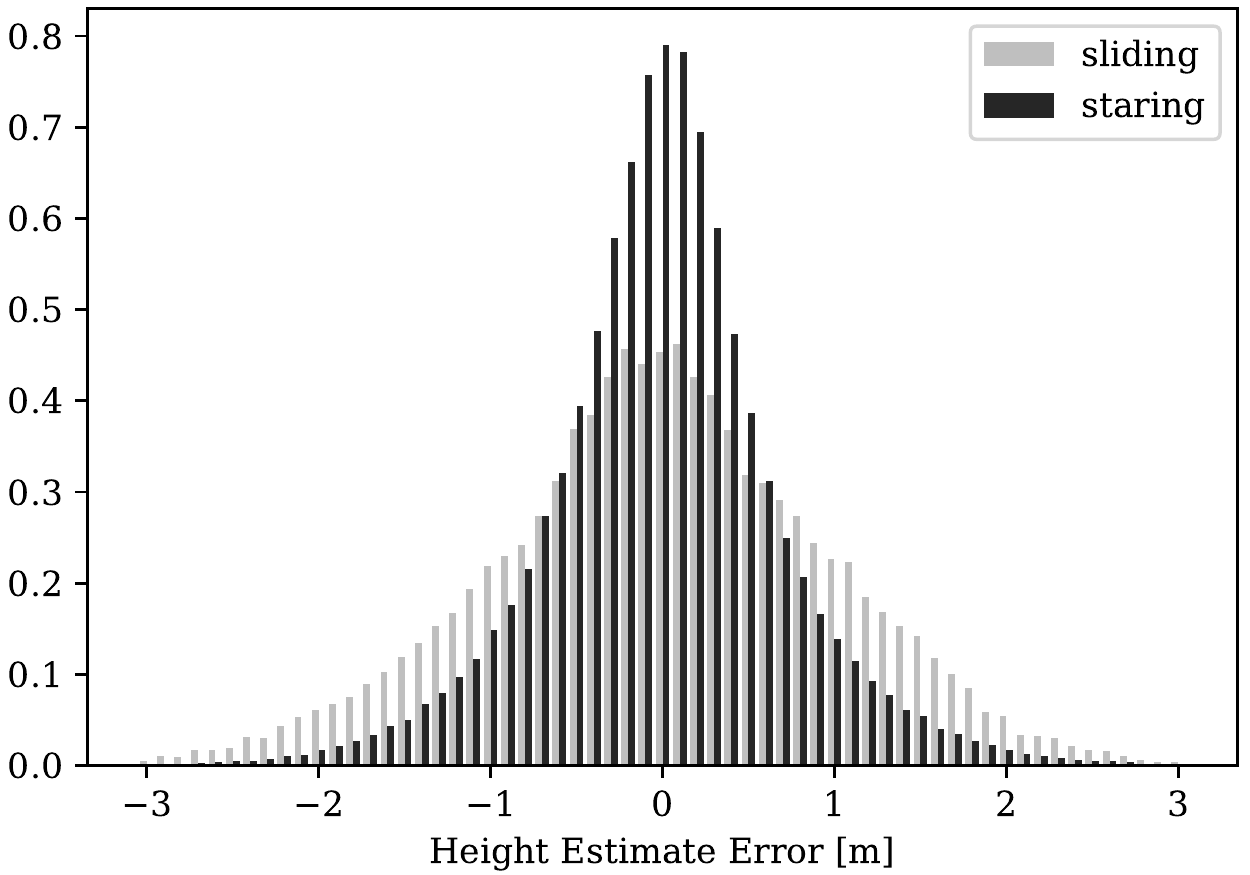}
\caption{Normalized histogram of height estimate errors of the point clouds in Fig.~\ref{fig:18} relative to a fitted plane.}
\label{fig:19}
\end{figure}

Fig.~\ref{fig:19} depicts the errors of height estimates relative to the fitted plane. Although both normalized histograms are centered around zero, the height estimate errors in the staring spotlight mode exhibit less deviation. According to Tab.~\ref{tab:3}, the relative height accuracy (defined as the median absolute deviation of height estimate errors) in the sliding spotlight case is approximately \(1.7\) times as high.

\begin{figure}[!tp]
\centering
\subfloat[Sliding]{\includegraphics[width=.25\textwidth]{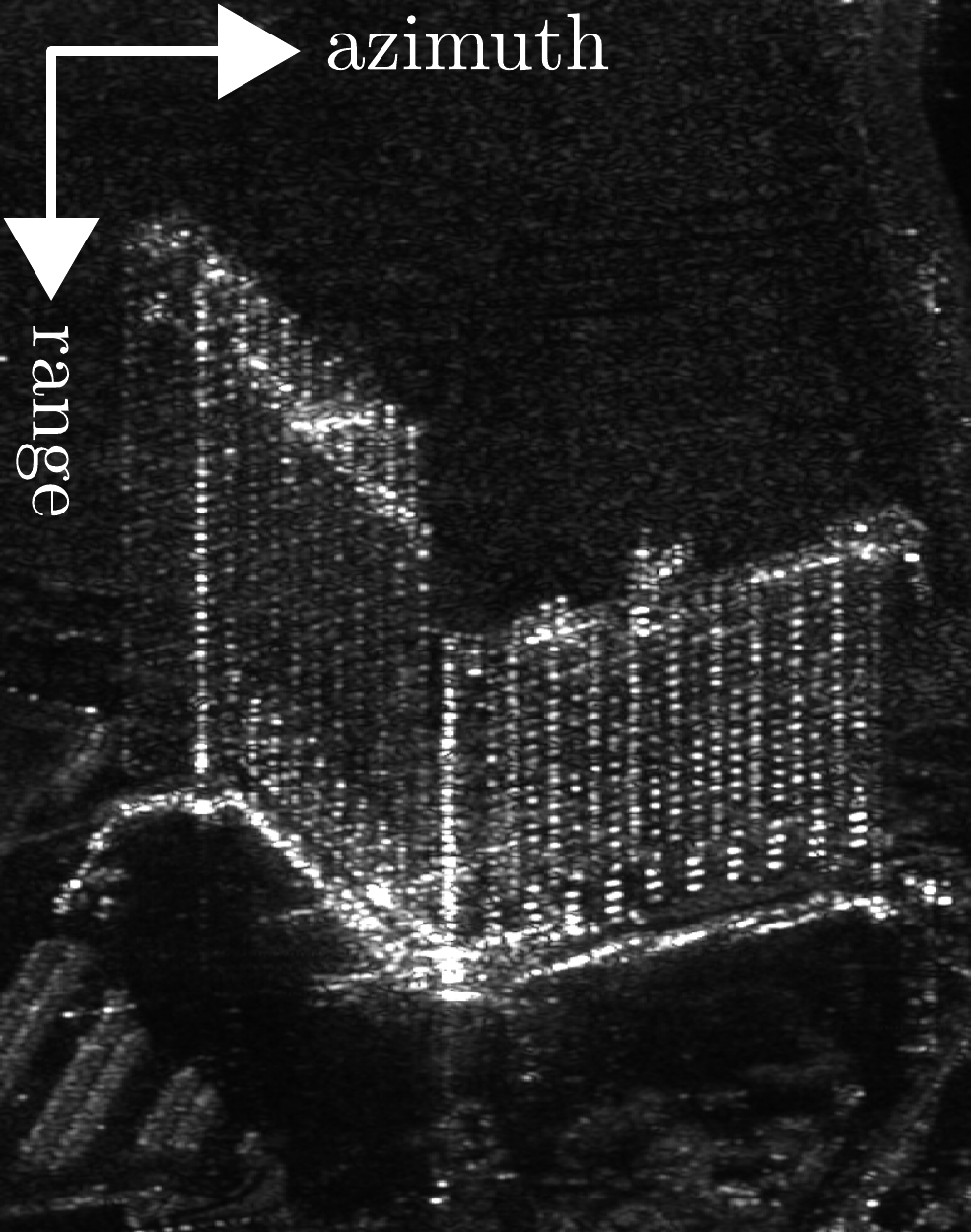}
\label{fig:20a}}
\subfloat[Staring]{\includegraphics[width=.25\textwidth]{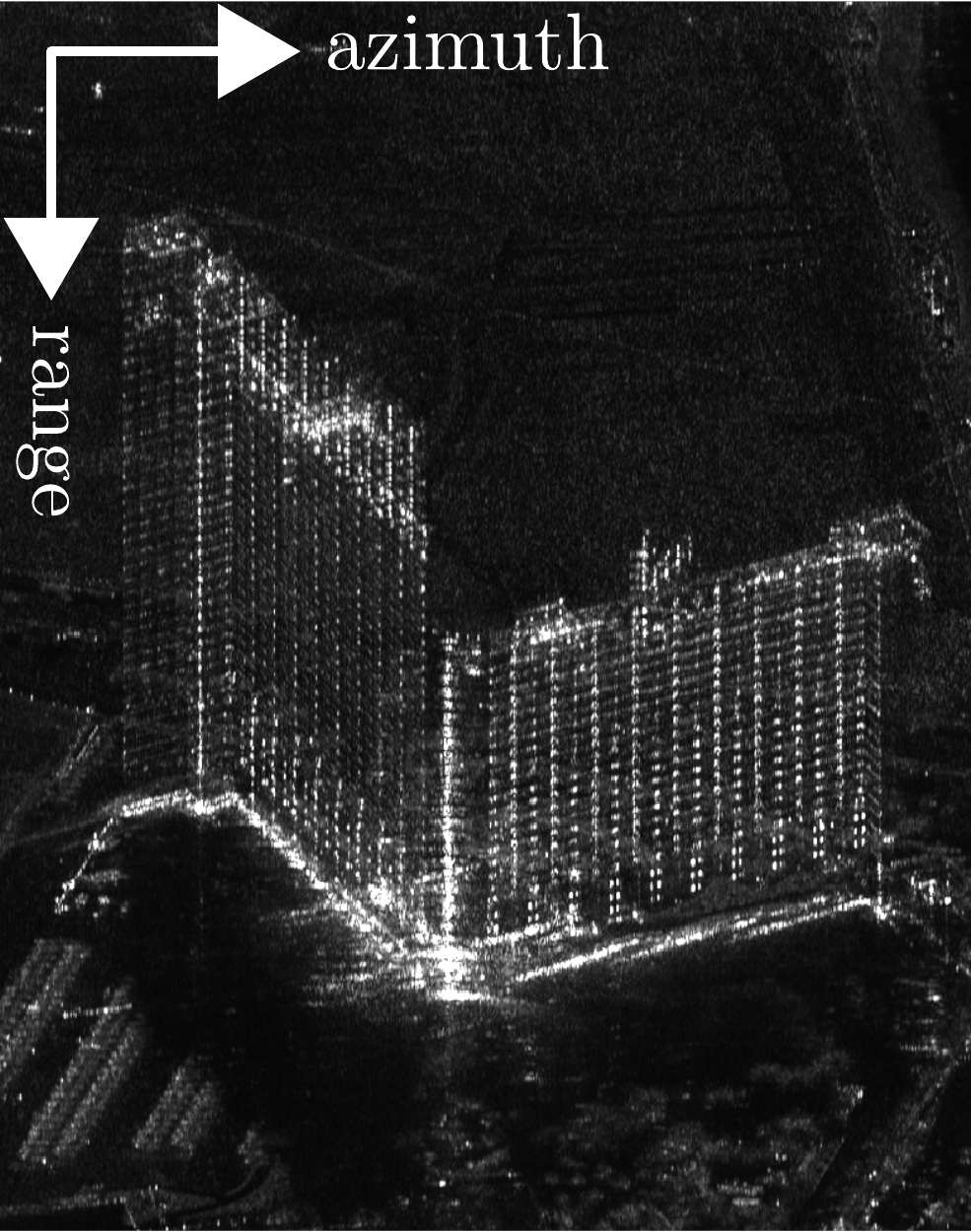}
\label{fig:20b}}
\caption{Mean intensity map of Hilton Grand Vacations on the Las Vegas Strip and its surroundings in the (a) sliding and (b) staring spotlight modes.}
\label{fig:20}
\end{figure}

\begin{figure*}[!tp]
\centering
\subfloat[Single scatterers (sliding)]{\includegraphics[width=.4\textwidth]{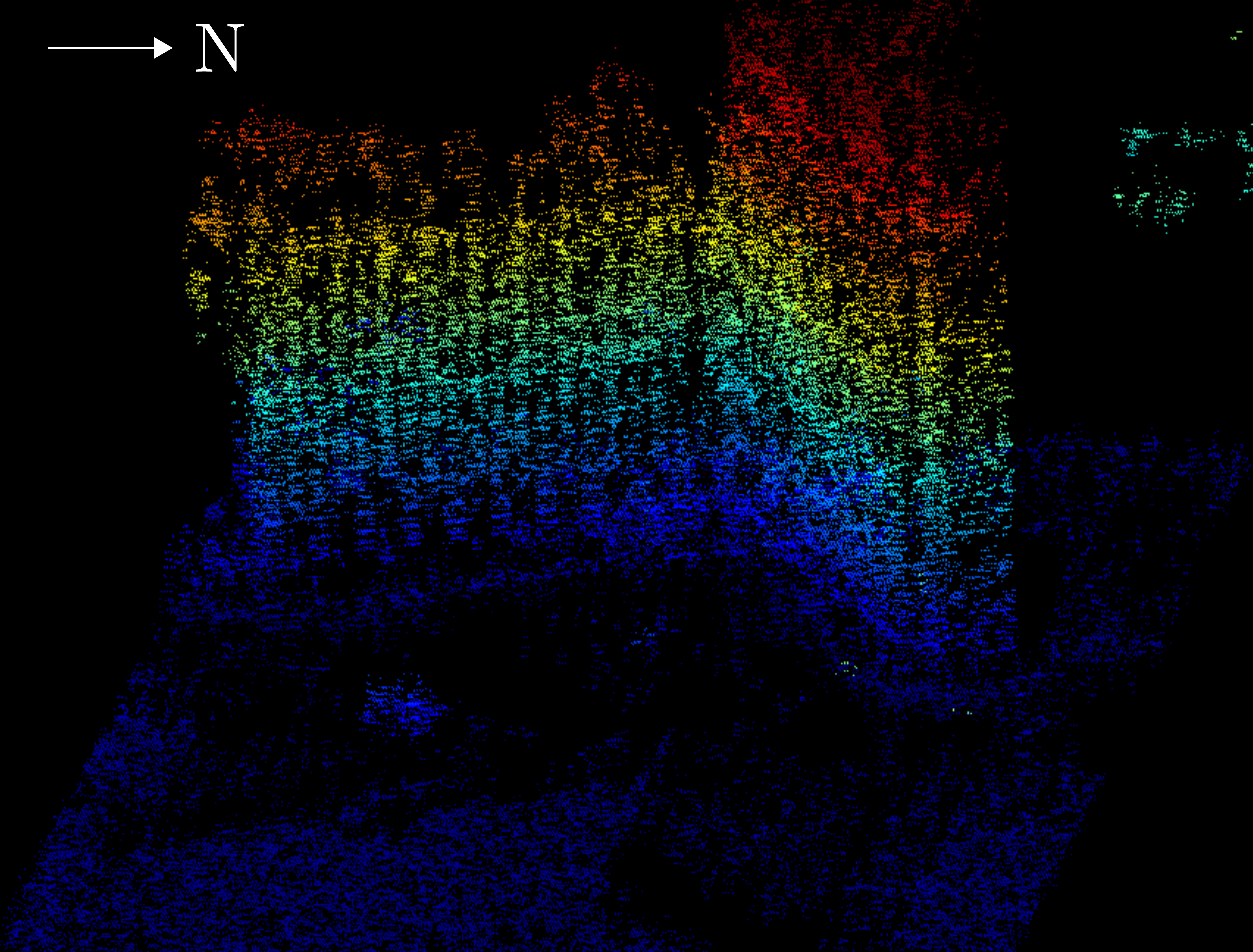}
\label{fig:21a}}
\subfloat[Single scatterers (staring)]{\includegraphics[width=.4\textwidth]{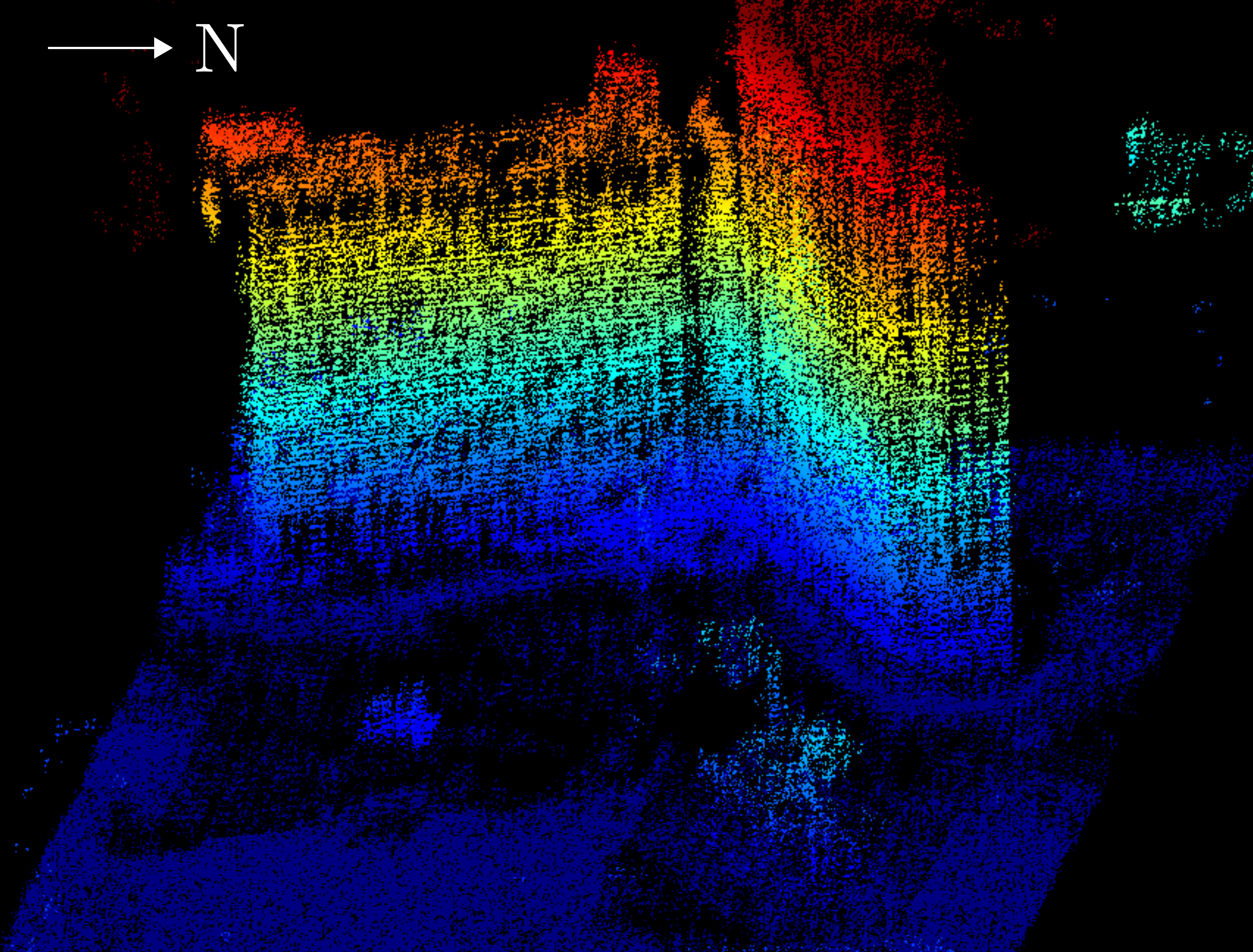}
\label{fig:21b}}
\hfil
\subfloat[Double scatterers (sliding)]{\includegraphics[width=.4\textwidth]{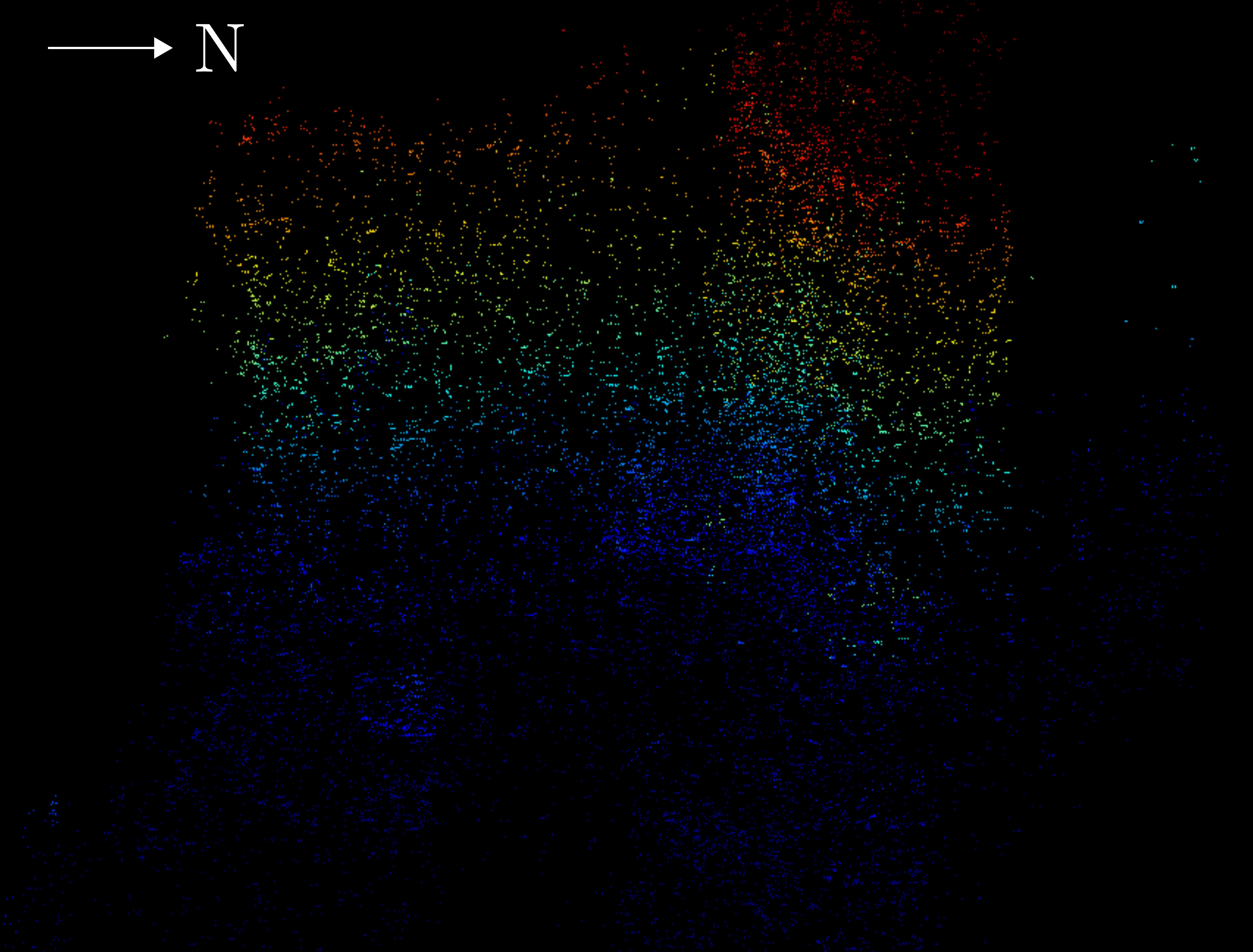}
\label{fig:21c}}
\subfloat[Double scatterers (staring)]{\includegraphics[width=.4\textwidth]{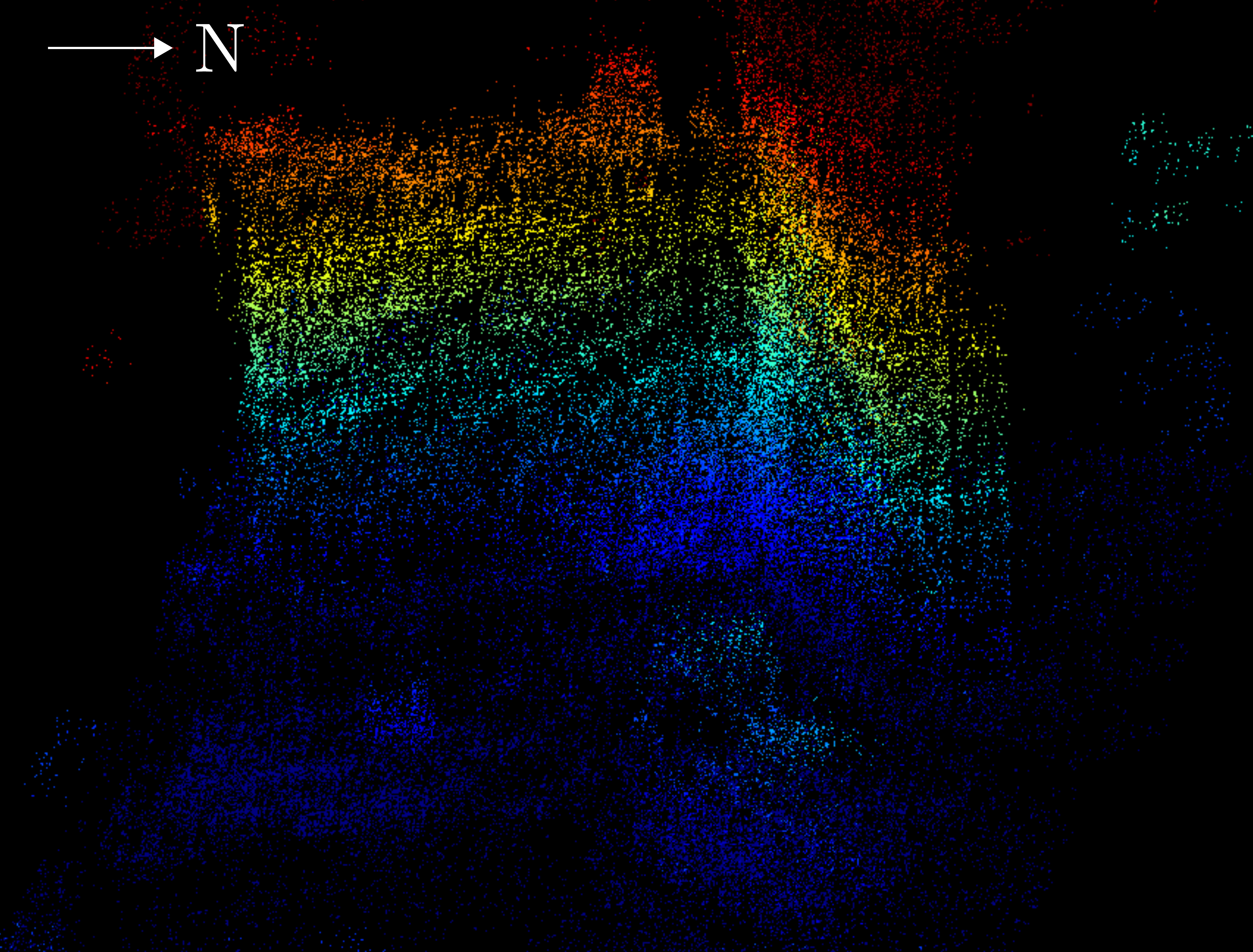}
\label{fig:21d}}
\hfil
\subfloat{\includegraphics[width=.3\textwidth]{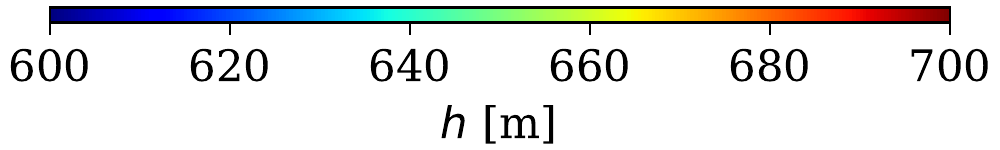}}
\caption{Updated topography \(h\) [m] of the area in Fig.~\ref{fig:20} with \(12\) TerraSAR-X images in the sliding (left column) and staring (right column) spotlight modes, respectively. The upper and lower rows show single and double scatterers, respectively.}
\label{fig:21}
\end{figure*}

The other area of approximately \(0.11\) km\(^2\) contains two high-rise buildings and its surroundings. The regular patterns of building facades appear sharper in the staring spotlight mode (see Fig.~\ref{fig:20}). The reconstructed point clouds are illustrated in Fig.~\ref{fig:21} for single and double scatterers, respectively. As expected, the staring spotlight mode densified the corresponding point cloud in both single- and double-scatterer cases. In total, the point density in the staring spotlight case is approximately \(5.1\) times as high, see Tab.~\ref{tab:4}. With respect to the ratio of the number of single scatterers to the number of double scatterers, we recorded a slight decrease approximately from \(6.9\) (sliding) to \(6.0\) (staring), i.e., no significant difference was observed.

\begin{table}[!tp]
    \renewcommand{\arraystretch}{1.3}
    \caption{Statistics of the point clouds in Fig.~\ref{fig:18}}
    \label{tab:2}
    \centering
    \begin{tabular}{l r r r}
        & Sliding & Staring & Ratio\footnotemark[2]\\
        \hline
        Total no.\ of scatterers & $26037$ & $142085$ & $5.46$\\
        Scatterer density [million/km\(^2\)] & $2.47$ & $13.46$ & $5.46$\\
        \hline
    \end{tabular}
\end{table}

\begin{table}[!tp]
    \renewcommand{\arraystretch}{1.3}
    \caption{Statistics of the height estimate errors in Fig.~\ref{fig:19}}
    \label{tab:3}
    \centering
    \begin{tabular}{l r r r}
        & Sliding & Staring & Ratio\footnotemark[2]\\
        \hline
        Median [m] & $0.00$ & $0.00$ & n.a.\\
        Mean [m] & $0.01$ & $0.01$ & n.a.\\
        Median absolute deviation [m] & $0.94$ & $0.54$ & $1.74$\\
        Standard deviation [m] & $1.12$ & $0.76$ & $1.47$\\
        \hline
    \end{tabular}
\end{table}

\begin{table}[!tp]
    \renewcommand{\arraystretch}{1.3}
    \caption{Statistics of the point clouds in Fig.~\ref{fig:21}}
    \label{tab:4}
    \centering
    \begin{tabular}{l r r r}
        & Sliding & Staring & Ratio\footnotemark[2]\\
        \hline
        No.\ of single scatterers & $148646$ & $740656$ & $4.98$\\
        No.\ of double scatterers & $21576$ & $124546$ & $5.77$\\
        Total no.\ of scatterers & $170222$ & $865202$ & $5.08$\\
        Single-to-double-scatterer ratio & $6.89$ & $5.95$ & $1.16$\\
        Scatterer density [million/km\(^2\)] & $1.56$ & $7.91$ & $5.08$\\
        \hline
    \end{tabular}
\end{table}

\footnotetext[2]{The ratio was calculated by dividing the larger by the smaller value.}

\section{Conclusion} \label{sec:6}
In this paper, we studied the characteristics of the TerraSAR-X staring spotlight mode and its impact on multibaseline InSAR techniques, in particular, PSI and TomoSAR. The difference in the time-variant Doppler spectra of the sliding and staring spotlight modes was analyzed in concept in order to demonstrate the azimuth resolution versus scene extent trade-off. The usage of the TerraSAR-X annotation component containing the Doppler centroid in focused image time was proposed to skirt the time conversion issue. The TomoSAR processing chain was revised in order to incorporate sidelobe detection, off-grid correction and outlier rejection. A first practical demonstration was made with an interferometric stack of $41$ images of Washington, D.C. The whole scene extent was processed to estimate topography update of point-like scatterers and their deformation parameters. Besides, the results of several typical urban areas were visualized and interpreted. A preliminary comparison between sliding and staring spotlight TomoSAR was drawn in the end with two small interferometric stacks of the City of Las Vegas.

In section~\ref{sec:1}, we argued that by means of the staring spotlight mode,
\begin{enumerate*}[label=\arabic*)]
\item more point-like targets would be separable in the azimuth-range plane;
\item each target would have a higher SCR.
\end{enumerate*}
As a result, the 4-D point cloud would be not only denser but also more accurate. In this work, we observed that,
\begin{enumerate*}[label=\arabic*)]
\item the density of the \emph{staring} spotlight point cloud is approximately \(5.1\)--\(5.5\) times as high;
\item the relative height accuracy of the \emph{staring} spotlight point cloud is approximately \(1.7\) times as high.
\end{enumerate*}

Multiple-snapshot TomoSAR approaches, e.g., using an adaptive neighborhood identified within a spatial search window \cite{for:15, sch:13}, or incorporating additional geospatial information of building footprints \cite{zhu:15}, could also benefit from the staring spotlight mode. In the former case, the enhanced azimuth resolution would increase the number of pixels in the homogeneous area; in the latter, the iso-height clusters of a facade to be jointly reconstructed would expand. On the whole, it would lead to a larger number of snapshots and in turn to a better estimation accuracy.

\begin{figure}[!tp]
\centering
\includegraphics[width=0.30\textwidth]{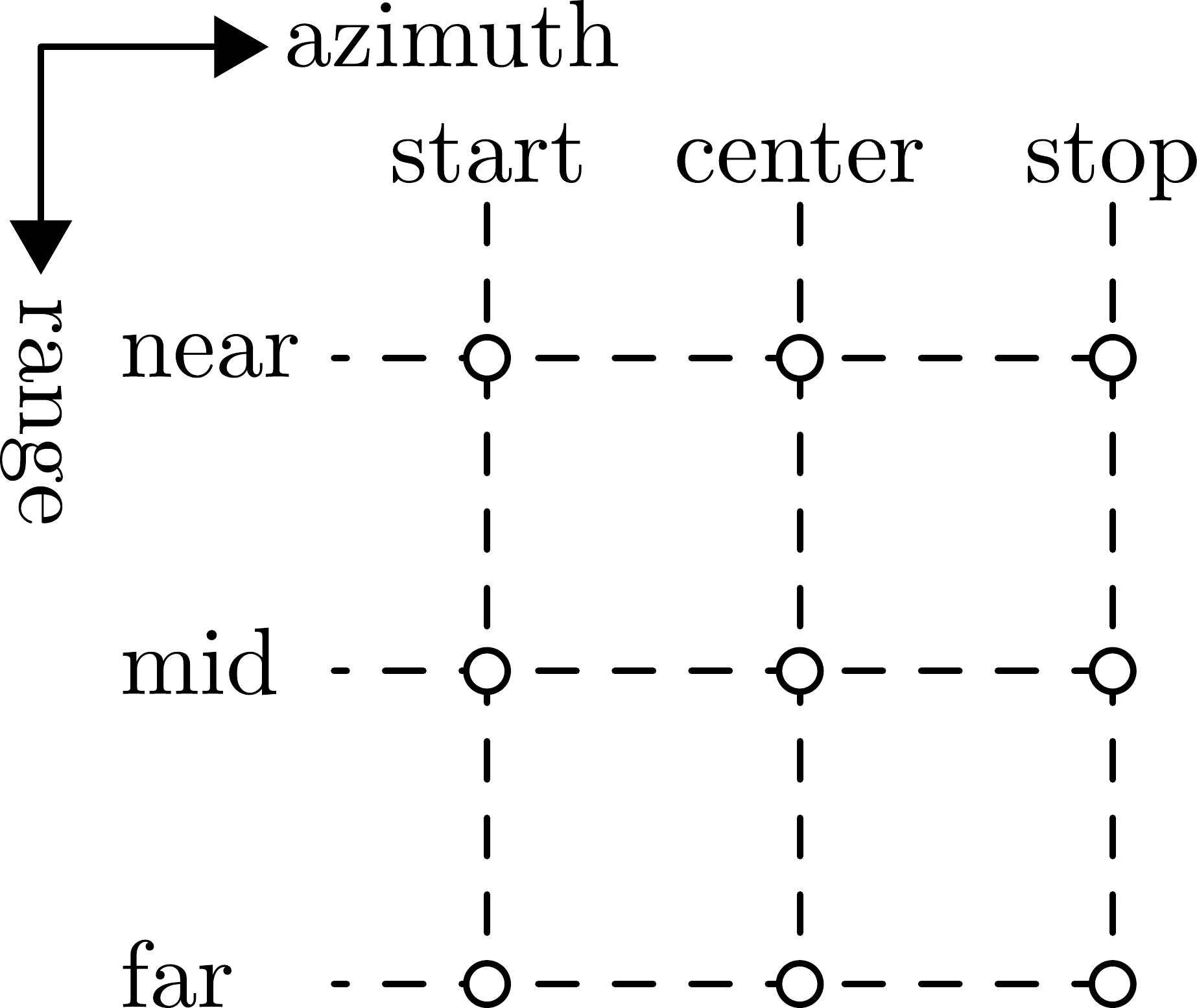}
\caption{$3$-by-$3$ grid of Doppler centroid frequency $f_{\text{DC}}$ in focused image time $t_{\text{image}}$.}
\label{fig:22}
\end{figure}

\appendix
As previously mentioned in section \ref{sec:2}, $f_{\text{DC}}$ is provided in $t_{\text{image}}$ on a $3$-by-$3$ grid as a TerraSAR-X annotation component \cite{fri:07}. This grid is defined as the Cartesian product of the sets $\{\text{start } t_{\text{image}}, \text{ center } t_{\text{image}}, \text{ stop } t_{\text{image}}\}$ and $\{\text{near range}, \text{ mid range}, \text{ far range}\}$, as depicted in Fig.~\ref{fig:22}. This information could be employed to bypass time conversion from $t_{\text{raw}}$ to $t_{\text{image}}$, and to consider second-order variations of $f_{\text{DC}}$ along range. Note that this grid is also provided for each burst of any ScanSAR SSC product.

%



\section*{Acknowledgment}
TerraSAR-X data was provided by the German Aerospace Center (DLR) under the TerraSAR-X New Modes AO Project LAN2188 and the TanDEM-X Science Phase AO Project NTI\_INSA6729. The authors would like to express their gratitude to TerraSAR-X science coordinator Ursula Marschalk for her kind support. The authors would also like to thank Helko Breit and Dr.~Thomas Fritz for their advice about TerraSAR-X annotation components, Nico Adam for his comment on the relation between spatial resolution and the density of double scatterers, Sina Montazeri for sharing his experience of sliding spotlight TomoSAR, Dr.~Marie Lachaise for the discussion about vertical accuracy, Alessandro Parizzi for explaining the principles of sidelobe detection, and the reviewers for their constructive and insightful comments.

\ifCLASSOPTIONcaptionsoff
  \newpage
\fi



\bibliographystyle{IEEEtran}
\bibliography{manuscript}

\end{document}